\documentclass[twocolumn]{aastex62}
\usepackage{graphicx}	
\usepackage{amsmath}	
\usepackage{amssymb}	

\received{August 15, 2019}
\accepted{November 26, 2019}

\shorttitle{The 1.28 GHz MeerKAT DEEP2 Image}
\shortauthors{Mauch et al.}

\begin{document}

\title{The 1.28 GHz MeerKAT DEEP2 Image}

\correspondingauthor{Tom Mauch}
\email{tmauch@ska.ac.za}

\author{T.~Mauch}
\affiliation{South African Radio Astronomy Observatory,
       2 Fir Street, Black River Park,
       Observatory 7925, South Africa}

\author{W.~D.~Cotton}
\affiliation{National Radio Astronomy Observatory,
             520 Edgemont Road,
             Charlottesville, VA 22903, USA}
\affiliation{South African Radio Astronomy Observatory,
             2 Fir Street, Black River Park,
             Observatory 7925, South Africa}

\author{J.~J.~Condon}
\affiliation{National Radio Astronomy Observatory,
             520 Edgemont Road,
             Charlottesville, VA 22903, USA}

\author{A.~M.~Matthews}
\affiliation{Department of Astronomy,
             University of Virginia,
             Charlottesville, VA 22904, USA}
\affiliation{National Radio Astronomy Observatory,
             520 Edgemont Road,
             Charlottesville, VA 22903, USA}

\author{T.~D.~Abbott}
\affiliation{South African Radio Astronomy Observatory, 2 Fir Street, Black River Park, Observatory 7925, South Africa}
\author{R.~M.~Adam}
\affiliation{South African Radio Astronomy Observatory, 2 Fir Street, Black River Park, Observatory 7925, South Africa}
\author{M.~A.~Aldera}
\affiliation{Tellumat (Pty) Ltd. 64-74 White Road, Retreat 7945, South Africa}
\author{K.~M.~B.~Asad}
\affiliation{South African Radio Astronomy Observatory, 2 Fir Street, Black River Park, Observatory 7925, South Africa}
\affiliation{Department of Physics and Electronics, Rhodes University, PO Box 94, Grahamstown 6140, South Africa}
\affiliation{Department of Physics and Astronomy, University of the Western Cape, Bellville, Cape Town 7535, South Africa}
\author{E.~F.~Bauermeister}
\affiliation{South African Radio Astronomy Observatory, 2 Fir Street, Black River Park, Observatory 7925, South Africa}
\author{T.~G.~H.~Bennett}
\affiliation{South African Radio Astronomy Observatory, 2 Fir Street, Black River Park, Observatory 7925, South Africa}
% ...

\begin{abstract}
We present the confusion-limited 1.28 GHz MeerKAT DEEP2 image
covering one $\theta_\mathrm{b} \approx 68\arcmin \mathrm{~FWHM}$
primary beam area with $\theta = 7\,\farcs6$~FWHM resolution and
$\sigma_\mathrm{n} = 0.55 \pm 0.01 \,\mathrm{{\mu}Jy\,beam^{-1}}$ rms
noise.  Its J2000 center position
$\alpha=04^\mathrm{h}\,13^\mathrm{m}\,26\,\fs4$,
$\delta=-80\degr\,00\arcmin\,00\arcsec$ was selected to minimize
artifacts caused by bright sources.  We introduce the new 64-element
MeerKAT array and describe commissioning observations to measure the
primary beam attenuation pattern, estimate telescope pointing errors,
and pinpoint $(u,v)$ coordinate errors caused by offsets in frequency
or time.  We constructed a 1.4\,GHz differential source count
by combining a power-law count fit to the DEEP2 confusion $P(D)$ distribution
from $0.25$ to $10\,\mu\mathrm{Jy}$ with counts of individual DEEP2 sources
between $10\,\mu\mathrm{Jy}$ and 2.5\,mJy.
Most sources fainter than $S \sim 100 \,\mu\mathrm{Jy}$ are distant
star-forming galaxies obeying the FIR/radio correlation, and sources
stronger than $0.25\,\mu\mathrm{Jy}$ account for $\sim93$\% of the
radio background produced by star-forming galaxies.  For the first
time, the DEEP2 source count has reached the depth needed to reveal
the majority of the star formation history of the universe. A
pure luminosity evolution of the
1.4\,GHz local luminosity function consistent with the
\citet{madaudickinson14} model for the evolution of star-forming
galaxies based on UV and infrared data underpredicts our 1.4\,GHz
source count in the range $-5 \lesssim \log[S\mathrm{(Jy)}] \lesssim
-4$.

\end{abstract}

\keywords{telescopes -- galaxies: statistics -- radio continuum:
  galaxies -- galaxies: star formation}



\section{Introduction}

The extragalactic source population at $1.4$\,GHz is a mixture of
galaxies with active galactic nuclei (AGNs) and star-forming galaxies
(SFGs) \citep{condon88,afonso05,simpson06,padovani15}.  Radio sources
powered by AGNs account for nearly all of the strong-source
population, and SFGs whose radio emission primarily comes from
synchrotron electrons accelerated by the supernova remnants of massive
($M>8M_\odot$) stars \citep{condon92} dominate below $S \sim 100
\,\mu\mathrm{Jy}$ at 1.4\,GHz \citep{simpson06,bonzini13,prandoni18}.

The far-infrared (FIR) / radio correlation obeyed by nearly all star-forming
galaxies indicates that their 1.4\,GHz luminosities are directly
proportional to their star-formation rates (SFRs) \citep{condon92}.
Dust is transparent at 1.4\,GHz, so sufficiently sensitive radio
continuum observations could trace the cosmic evolution of the mean
star-formation rate density (SFRD) unbiased by dust emission or
absorption.

In the past two decades a number of wide-area redshift surveys
\citep{colless01,york00,jones09} used in combination with
mJy-sensitivity radio surveys such as the NRAO VLA Sky Survey
\citep[NVSS;][]{condon98} and the Sydney University Molonglo Sky
Survey \citep[SUMSS;][]{mauch03} have allowed accurate determinations
of the local radio luminosity functions (RLFs) for both SFGs and
AGNs \citep{sadler02,best05,mauch07,condon19}. In all cases the two
populations were classified using available optical, mid-infrared,
FIR, and radio data.

Measuring the evolving SFRD by directly determining RLFs at higher
redshifts and comparing them with the well-determined local RLF is
difficult.  It requires deep multi-wavelength data to identify and
classify the host galaxies of faint radio sources, plus photometric or
spectroscopic redshifts.  Recent studies of SFRD evolution by this
method \citep[e.g.][]{smolcic09,padovani11,mcalpine13} suggest pure
luminosity evolution $\propto(1+z)^{2.5}$ for the most luminous SFGs
at $z \lesssim 2.5$.  However, current radio surveys are not
sensitive enough to detect the fainter galaxies responsible for the
bulk of star formation around ``cosmic noon'' at $z \sim 2$. Most
current samples are further hampered by their reliance on deep
multi-wavelength data covering very small solid angles in fields
selected using only optical/infrared criteria, which can be less than
ideal for making deep radio images.
A $5\sigma_\mathrm{n} \approx0.25\,\mu$Jy\,beam$^{-1}$ sensitivity is
needed to detect the individual SFGs accounting for most of the star
formation history of the universe (SFHU), so making a survey with rms
noise $\sigma_\mathrm{n} \approx 0.05 \,\mu\mathrm{Jy\,beam}^{-1}$ is
one of the key continuum science goals of the Square Kilometre Array
(SKA) \citep{jarvis15,prandoni15}.

Confusion by sources blended in the synthesized beams of deep
continuum images also limits the ability of radio surveys to detect
faint SFGs. For example, the MeerKAT MIGHTEE survey will have
rms confusion $\sigma_\mathrm{c} \approx 2\,\mu$Jy\,beam$^{-1}$ in its
$\theta \approx8\arcsec$ synthesized beam \citep{jarvis16}.  The
detection threshold for individual sources will be about
$17\,\mu\mathrm{Jy}$, the level at which there are $\sim 25$ beam solid
angles per source and below which the association of fitted components
with individual galaxies is hampered by the increasing level of
obscuration by stronger ones.  Recent work by \citet{condon12} and
\citet{vernstrom14} has shown that statistical analysis of the
confusion brightness distribution, usually called the $P(D)$
distribution, can be used to estimate the source count at sub-$\mu$Jy
levels. Such very deep source counts combined with the already
accurately determined local RLF can be used to constrain the SFHU
directly \citep{condon18}.

This paper presents the 1.28 GHz DEEP2 image observed with the South
African Radio Astronomy Observatory's (SARAO) MeerKAT array,
counts of radio sources in DEEP2, and a preliminary
analysis of the SFHU constrained by those counts.
Section~\ref{sec:meerkat} provides a brief introduction to the MeerKAT
array and describes our early commissioning observations designed to
measure and improve its performance.  In order to maximize the depth
we can reach with the finite dynamic range of the MeerKAT
array, we chose the DEEP2 field to be as free as possible from bright
radio sources in the MeerKAT primary beam, as described in
Section~\ref{sec:selection}. Our observing strategy for the DEEP2
field plus our method of calibration and imaging the raw data are
outlined in Section~\ref{sec:obsim}.  Section~\ref{sec:counts}
presents source counts from 0.25 to $10\,\mu\mathrm{Jy}$
derived from the DEEP2 image $P(D)$ distribution and between $10\,\mu\mathrm{Jy}$
and 2.5~mJy based on discrete sources in DEEP2.
Finally, for Section~\ref{sec:sfhu} we evolved the local
1.4~GHz luminosity function of star-forming galaxies \citep{condon19}
for pure luminosity evolution of the \citet{madaudickinson14} fit to
the UV/FIR SFHU and compared it with our 1.4\,GHz source count.

\section{The MeerKAT Radio Telescope} \label{sec:meerkat}

The MeerKAT array was used to observe the DEEP2 field.
MeerKAT is a precursor to the Square Kilometre Array (SKA) located in
the Karoo region of South Africa's Northern Cape province. MeerKAT was
inaugurated in July 2018 and, during the course of our DEEP2 observations,
was in its early commissioning phase. MeerKAT is a new instrument
that has little presence in the literature, so we describe the telescope features
(Section~\ref{subsec:MeerKAT}) and results from our early
commissioning observations (Section~\ref{subsec:1934})
relevant for deep continuum observations.

\subsection{MeerKAT} \label{subsec:MeerKAT}

MeerKAT is an array of 64 13.5\,m-diameter dish antennas spread out over roughly
8\,km centered near latitude $30\degr\,42\arcmin$\,S and longitude
$21\degr\,23\arcmin$\,E. Each dish has a 3.8\,m offset Gregorian
subreflector and a receiver indexer located just below the subreflector to
ensure a completely unblocked aperture. A conical skirt extends below the
subreflector to deflect radiation from the surrounding ground away from the
receiver. The unblocked aperture is essential for deep continuum imaging at L band
because it improves dynamic range by (1) lowering the sensitivity of
primary beam sidelobes to strong sources and RFI outside the main beam and
(2) reducing the system noise temperature by limiting pickup of ground radiation.
The observations described in this paper were all carried out with the dual
linear polarization (horizontal and vertical) L-band (856--1712~MHz) receivers
\citep{lehmensiek12,lehmensiek14}.  Each antenna has a measured system noise
temperature $T_\mathrm{sys} \approx 20$~K and a remarkably low system equivalent
flux density $\mathrm{SEFD} \approx 430\,\mathrm{Jy}$ on cold sky.

\begin{figure}
  \includegraphics[width=\columnwidth]{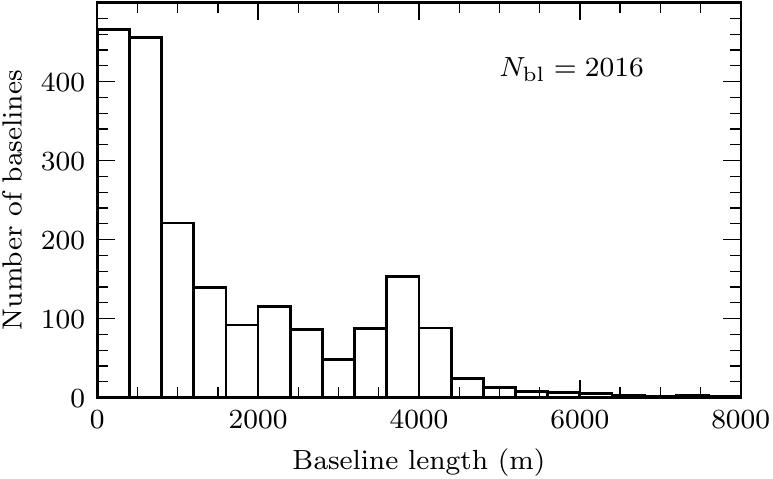}
    \caption{The distribution of all $N_\mathrm{bl}=2016$ MeerKAT baseline
             lengths which range between 29 and 7698 m.
             Half of the baselines are between the 48 antennas in
             the densely packed 1\,km diameter core.}
    \label{fig:mkat_blines}
\end{figure}

\begin{figure}
  \includegraphics[width=\columnwidth]{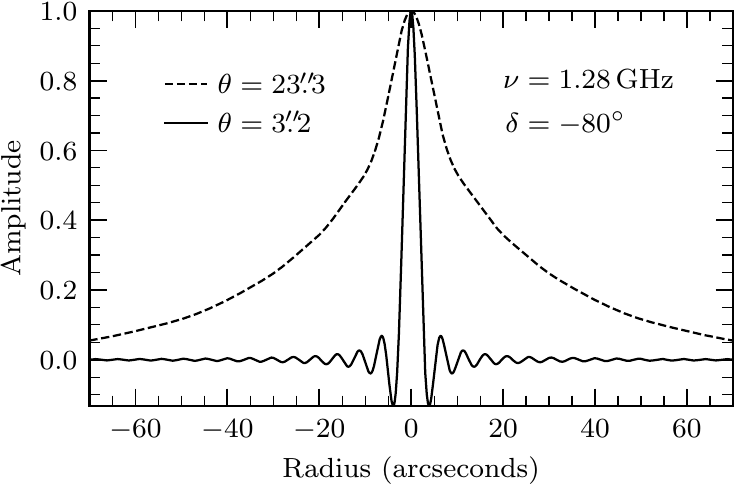}
    \caption{Two slices through synthesised beams derived from a
             simulated 12 hour MeerKAT observation at ${\delta = -80^\circ}$.
             The dashed line shows a beam calculated using natural weights
             (i.e. constant per visibility), and has FWHM
             ${\theta = 23\,\farcs3}$. The solid line shows a beam calculated
             with uniform weights (i.e. inversely proportional to the number of
             visibilities in a grid sample), and has FWHM
             $\theta = 3\,\farcs2$.}
    \label{fig:beam_slice}
\end{figure}

The MeerKAT antennas are named ``m000'' through ``m063'', the order
of which roughly follows their distances from the array center. The
inner 48 antennas are located within a 1\,km diameter central ``core'' region,
and the remaining 16 are spread beyond this central area out to a
radius of nearly 4\,km. The distribution of baseline lengths between
the 2016 antenna pairs (Figure~\ref{fig:mkat_blines}) is peaked at
lengths shorter than 1000\,m, and roughly half of all MeerKAT
baselines are between antennas in the core.
MeerKAT and large proposed arrays such as the SKA and ngVLA have
fixed antennas in centrally concentrated multiscale configurations,
which are compromises designed to satisfy conflicting demands for
high surface-brightness sensitivity and high angular resolution.
MeerKAT provides excellent L band surface-brightness sensitivity on
angular scales $\gtrsim 1\arcmin$ at the cost of a very broad
naturally weighted synthesized beam
(Figure~\ref{fig:beam_slice}). To achieve $ \theta \approx
7\,\farcs6$ FWHM resolution in our DEEP2 image, we observed only
when most of the outer antennas were working, and we gave up
some sensitivity by heavily downweighting the $(u,v)$ data from
intra-core baselines.

Visibilities are transported to the archive located in the
Centre for High Performance Computing (CHPC) 600\,km away in Cape
Town, where they are converted
to formats used by major radio-astronomy imaging and analysis
packages (e.g., Measurement Set or AIPS UV).
Additional information about MeerKAT and its specifications
can be found in \citet{jonas16,camilo18}.

\subsection{Commissioning observations of PKS\,B1934$-$638} \label{subsec:1934}

To verify the accuracy of our early MeerKAT data we made a series
of snapshot images offset by 10, 20, 30, 40 and 50\,arcmin to the
north, south, east, and west of the calibration source
PKS\,B1934$-$638. At the $\theta \approx 8\arcsec$ resolution of
MeerKAT, PKS\,B1934$-$638 is a point source. It is also strong ($S
\approx15.1$\,Jy) at 1284\,MHz and has a well-established radio
spectrum \citep{reynolds94}. The usefulness of such offset snapshots
is described in \citet{condon98}: position errors can reveal incorrect
$(u,v)$ coordinates or frequency labeling in the data, and the
variation in amplitude with position in the primary beam can be used
to measure the primary-beam attenuation patterns and pointing errors
of the MeerKAT dishes.

A 14\,minute observation cycling through the offset pointings was made with a
start time chosen to ensure that the scans were as close to the
transit of PKS\,B1934$-$638 as possible (azimuth $\approx 180^\circ$,
elevation $\approx 57^\circ$). Images made from individual pointings have rms
noise $\sigma_\mathrm{n} \sim 1 \mathrm{\,mJy\,beam}^{-1}$. The flux
density and position of the source was measured in each image by
fitting an elliptical Gaussian using the AIPS task JMFIT.  At all
offsets the source is never attenuated to $S_\mathrm{a} <
2.3$\,Jy\,beam$^{-1}$ by the MeerKAT primary beam, so the source
signal-to-noise ratio is always $\mathrm{SNR} \gtrsim 2300$:1. The
noise component of fitting errors should be $< 0\,\farcs006$ in
position and $<0.04$\% in flux density assuming a point source and a
circular $8\arcsec$ beam \citep{condon97}, so errors in the
measured positions and flux densities are dominated by calibration
errors.

\subsubsection{Positions}

\begin{figure}
  \includegraphics[width=\columnwidth]{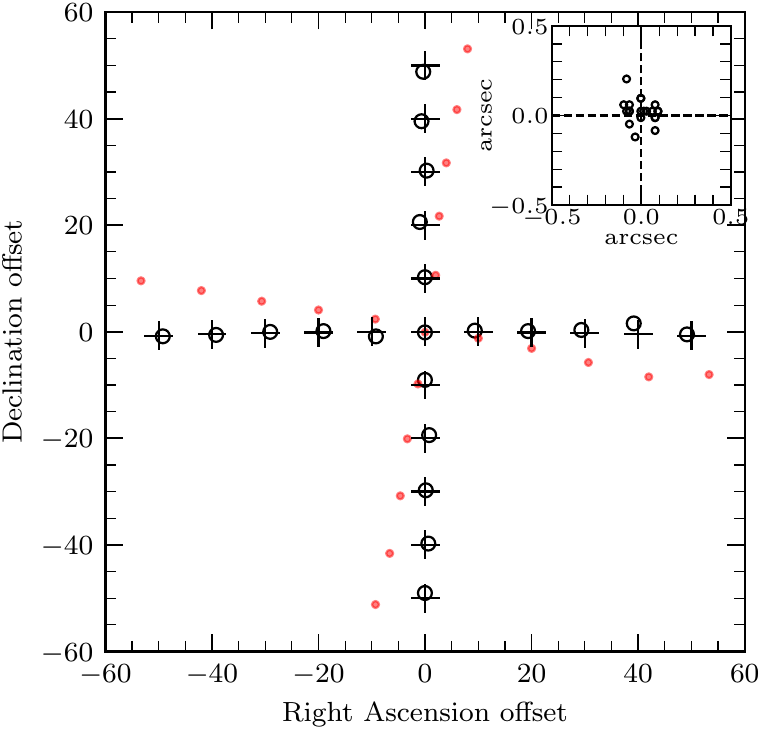}
  \caption{The commanded position offsets (arcmin) for the
    observations of PKS\,B1934$-$638 described in Section~\ref{subsec:1934}
    are shown as crosses.  The red points show the measured positions
    calculated from data with
    frequency and time errors; the open circles show the measured positions
    after correction for these errors.
    The differences between the measured and commanded
    offsets as shown have been magnified $600\times$ to highlight
    the small errors in the measured positions. The panel in the upper
    right shows the distribution of corrected right ascension and declination
    errors in units of arcsec for all 21 pointings.}
  \label{fig:crossoff}
\end{figure}

Geometric errors can be introduced into source positions on the image
plane by incorrect calculations of the $(u,v)$ coordinates associated
with measured visibilities. The $(u,v)$ coordinates used during imaging
were calculated at a reference frequency of $\nu_0=886$\,MHz.
If $\nu_0$ differs from the actual array reference frequency
$\nu_\mathrm{c}$, this will rescale the image radial offsets of
source positions from the phase center $\rho_\mathrm{m}$ from their true
offsets $\rho$ \citep{condon98}.  Furthermore, if the timestamps used to
calculate the $(u,v)$ coordinates differ from the true timestamps of
observation by $\Delta t$\,s, an image near the celestial pole will be rotated
about its phase center by an angle $\theta \approx 2 \pi \Delta t /
T_\mathrm{d}$ where $T_\mathrm{d} \approx 86164$\,s is one sidereal
day.

Figure~\ref{fig:crossoff} compares the measured offsets (circles and red points)
of the 21 pointings with their commanded offsets (crosses). The red points
indicate the measured offsets in our initial dataset before any
corrections; they clearly show both a rotation and a radial scaling. The rotation
revealed a 2 second shift in the \mbox{timestamps} of the raw correlated
visibilities. The radial scaling was caused by
the frequency labeling being offset by 0.5 channels during conversion of the
data to AIPS UV. Correcting these errors moved the red points
to the black circles in Figure~\ref{fig:crossoff}. To ensure that all MeerKAT data
do not require these time and frequency corrections, the errors have now been
fixed in the MeerKAT correlator and in the software.

The mean ratio of the corrected measured (circles) to the commanded (crosses)
radial offsets is consistent with unity:
\begin{equation}
  \langle \rho_\mathrm{m}/\rho \rangle
  = \nu_{c}/\nu_{o} = 1 - (4.0 \pm 4.9) \times10^{-5}~.
\end{equation}
Likewise, the circles in  Figure~\ref{fig:crossoff} are consistent with no
rotation about the center.

The upper right inset of Figure~\ref{fig:crossoff} shows the
distribution of position errors in right ascension and declination.
Their standard errors and means are
\begin{equation}
\begin{aligned}
\sigma_{\Delta\alpha} &= 0\,\farcs11,\qquad & \langle\Delta\alpha\rangle &= 0\,\farcs001, \\
\sigma_{\Delta\delta} &= 0\,\farcs06,\qquad & \langle\Delta\delta\rangle &= 0\,\farcs02.
\end{aligned}
\label{eqn:posnacc}
\end{equation}
Thus we expect that individual strong sources within $\sim 50\arcmin$
of the phase centers of L-band MeerKAT images will have rms position errors
$\sim 0\,\farcs1$ in each coordinate, and the image frames will have
astrometric uncertainties  $\sim 0\,\farcs01$.

\subsubsection{The MeerKAT Primary Beam}
\label{subsubsec:primarybeam}

\begin{figure}
  \includegraphics[width=\columnwidth]{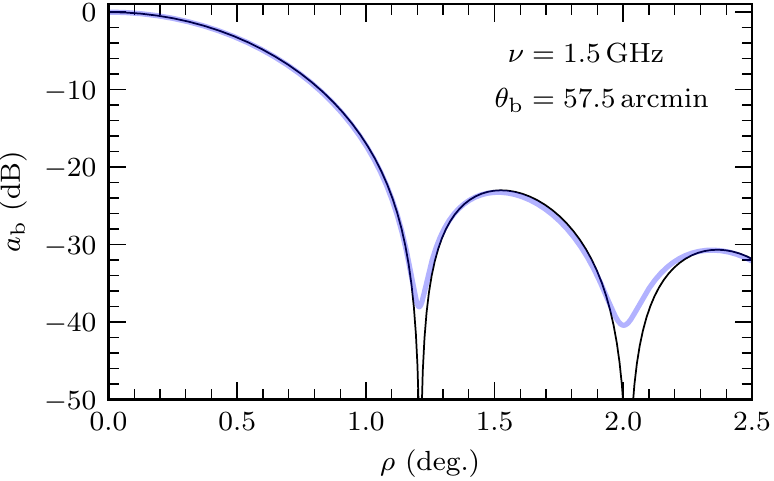}
  \caption{The blue curve shows a horizontal slice through
    holographic measurements of the 1.5\,GHz Stokes~I primary beam
    power pattern of MeerKAT (M.~de Villiers, in prep).
    The black curve is the attenuation pattern calculated from
    Equation~\ref{eqn:cospb} for FWHM $\theta_\mathrm{b} = 57\farcm5$.}
  \label{fig:pbfunc}
\end{figure}

The attenuated flux densities of PKS\,B1934$-$638, observed at various
pointing offsets $\rho$, divided by its flux density at the pointing
center, measure the primary beam attenuation pattern
 $a_\mathrm{b}(\rho) \equiv S(\rho)/S(\rho=0)$
of the MeerKAT antennas.
The attenuation pattern derived from these data
and also from  a horizontal slice through holographic measurements of the
MeerKAT Stokes I beam at 1.5\,GHz (M.~de Villiers, in prep) is well matched by
the attenuation pattern resulting from  cosine-tapered field (or cosine-squared power)
illumination \citep{condon16}:
    \begin{equation}
      a_\mathrm{b}(\rho/\theta_\mathrm{b}) =
      \left[\frac{\mathrm{cos}(1.189\pi\rho/\theta_\mathrm{b})}
      {1 - 4(1.189\rho/\theta_\mathrm{b})^2}\right]^2~.
      \label{eqn:cospb}
    \end{equation}
For the purpose of comparing the observed
attenuation pattern at 1.5\,GHz with Equation~\ref{eqn:cospb} we set
the FWHM of the horizontal slice through the primary power pattern to
\begin{equation}
  \theta_\mathrm{b} =  57\,\farcm5 \biggl( \frac {\nu}
  {1.5 \mathrm{\,GHz}} \biggr)^{-1} ~.
\end{equation}

Figure~\ref{fig:pbfunc} shows that the attenuation
pattern of the cosine illumination taper (Equation~\ref{eqn:cospb})
is a good match out to $\rho = 2\,\fdg5$.
Equation~\ref{eqn:cospb} also has the practical advantage of
expressing $a_\mathrm{b}(\rho/\theta_\mathrm{b})$ as an elementary function, so
we used it to fit our PKS\,B1934$-$638 data and in all subsequent
analyses requiring narrowband beam patterns at frequency $\nu$.

\begin{figure}
  \includegraphics[width=\columnwidth]{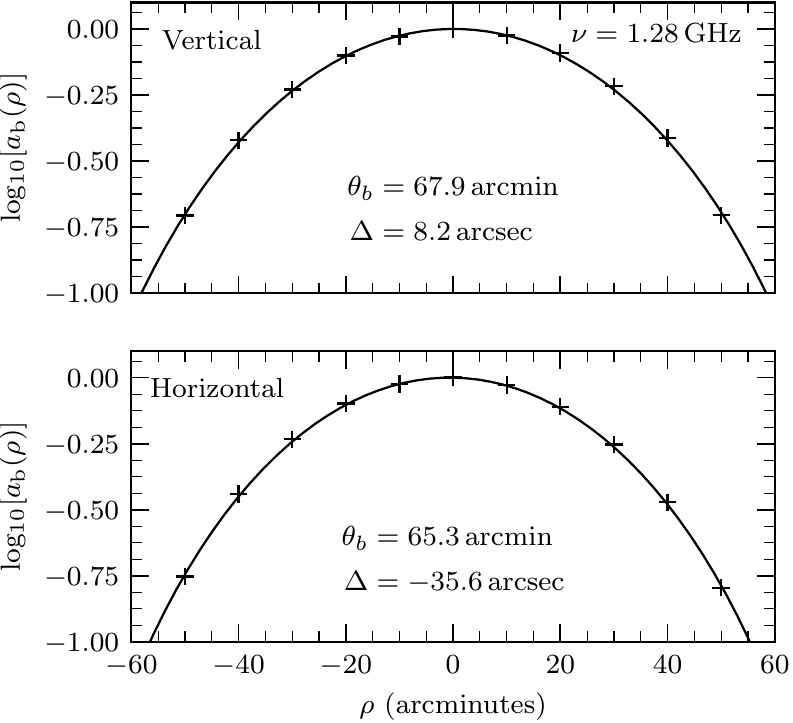}
    \caption{The fitted attenuation patterns of
      Equation~\ref{eqn:cospb} to vertical (top panel) and horizontal
      (bottom panel) slices through the Stokes I MeerKAT L-band
      primary beam. Crosses show the measured attenuation in the flux
      density $S$ of PKS\,B1934$-$638 derived from elliptical Gaussian
      fits to the source at each offset position from our
      observations. $\theta_\mathrm{b}$ denotes the best-fitting FWHMs
      at $\nu = 1.28 \mathrm{\,GHz}$ and $\Delta$ denotes the offsets
      of the peak in the fitted attenuation patterns from $\rho=0$.}
    \label{fig:cospb}
\end{figure}

Figure~\ref{fig:cospb} shows fits of Equation~\ref{eqn:cospb} to the
flux densities measured from our PKS\,B1934$-$638 observations.  During
fitting we inserted the mean pointing offset $\Delta$ as a free
parameter, replacing $(\rho/\theta_\mathrm{b})$ in
Equation~\ref{eqn:cospb} by $[(\rho -\Delta)/ \theta_\mathrm{b}]$.
The horizontal and vertical fits of the Equation~\ref{eqn:cospb}
attenuation power pattern are an excellent match to the data for all
$\log_{10}(a_\mathrm{b}) > -0.70$.

The primary beam pattern is slightly elliptical; its vertical
$\theta_\mathrm{b}$ is larger than its horizontal $\theta_\mathrm{b}$.
Simulations of the MeerKAT antenna optics predict that asymmetries in the
horizontal and vertical linearly polarized L-band feeds result in an
ellipticity in the Stokes~I primary beam pattern. Our measurement of the
ellipticity as a function of frequency (Table~\ref{tab:pbfreq}) agrees with
these simulations to within 1\%. We conclude that feed asymmetry is the
reason for the ellipticity in the measured Stokes I MeerKAT L-band primary beam.

\begin{table}
  \centering
  \caption{Frequency dependence of the MeerKAT primary beamwidths
           $\theta_\mathrm{b}$ and typical pointing errors $\Delta$}
  \label{tab:pbfreq}
  \begin{tabular}{crcrcc}
  \hline
  & & \multicolumn{2}{c}{Vertical} & \multicolumn{2}{c}{Horizontal} \\
  \multicolumn{1}{c}{Subban\rlap{d}} & \multicolumn{1}{c}{$\nu$} &
  $\theta_\mathrm{b}$ & \multicolumn{1}{c}{$\Delta$} &
  $\theta_\mathrm{b}$ & \multicolumn{1}{c}{$\Delta$} \\
  \multicolumn{1}{c}{numbe\rlap{r}} & \multicolumn{1}{c}{(MHz)} &
  \llap{(}arcmin\rlap{)}& \multicolumn{1}{c}{(arcsec\rlap{)}}& (arcmin) &
  \multicolumn{1}{c}{(arcsec)} \\
  \hline
1 & $908.04$ & $\llap{1}00.1$ & $-17.1$ & $96.2$ & $-21.7$ \\
2 & $952.34$ & $94.7$ & $-5.1$ & $91.1$ & $-24.1$ \\
3 & $996.65$ & $90.5$ & $-2.4$ & $87.1$ & $-23.4$ \\
4 & $1043.46$ & $86.1$ & $6.2$ & $82.8$ & $-25.4$ \\
5 & $1092.78$ & $81.7$ & $14.0$ & $78.6$ & $-27.8$ \\
6 & $1144.61$ & $78.2$ & $17.4$ & $75.2$ & $-27.0$ \\
7 & $1198.94$ & $73.4$ & $21.8$ & $70.5$ & $-32.2$ \\
8 & $1255.79$ & $70.0$ & $21.1$ & $67.3$ & $-29.5$ \\
9 & $1317.23$ & $65.7$ & $19.7$ & $63.2$ & $-25.9$ \\
\llap{1}0 & $1381.18$ & $63.1$ & $12.1$ & $60.6$ & $-32.6$ \\
\llap{1}1 & $1448.05$ & $60.6$ & $-29.7$ & $58.3$ & $-68.3$ \\
\llap{1}2 & $1519.94$ & $58.9$ & $-30.2$ & $56.8$ & $-63.6$ \\
\llap{1}3 & $1593.92$ & $56.2$ & $-19.4$ & $54.3$ & $-39.4$ \\
\llap{1}4 & $1656.20$ & $55.4$ & $-50.4$ & $53.6$ & $-43.2$ \\
  \hline
  \end{tabular}
\end{table}

We used wideband images of PKS B1934$-$638 covering the frequency
range 886--1682\,MHz centered on $\nu=1.28$\,GHz to fit the primary
beam in Figure~\ref{fig:cospb}. We have further split the band into 14
narrow subbands and repeated the beam fitting described above for each.
Table~\ref{tab:pbfreq} summarizes the results.  Note
that the narrow subbands defined in this table are the same as the
subbands in Table~\ref{tab:subbands} used for the wideband imaging of
the DEEP2 field.

\begin{figure}
  \includegraphics[width=\columnwidth]{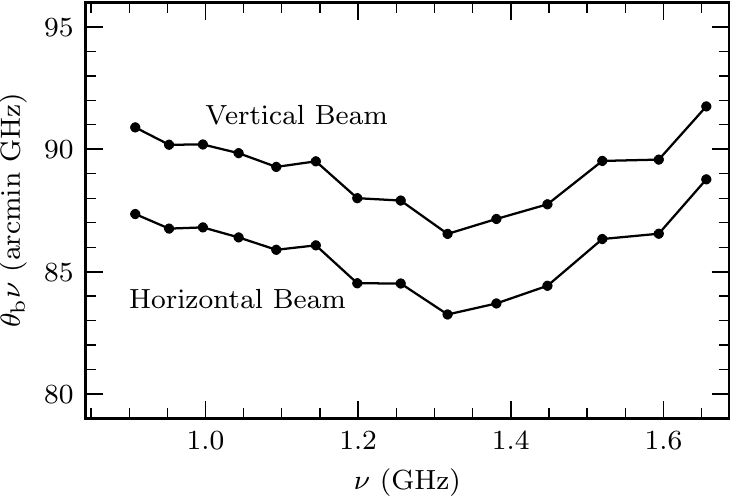}
  \caption{The variation of $\theta_\mathrm{b}\nu$ with frequency $\nu$ in
           GHz from the values tabulated in Table~\ref{tab:pbfreq}. The lower
           line shows values from fits of Equation~\ref{eqn:cospb} to a
           horizontal slice through the primary beam and the upper line shows
           fits to a vertical slice.}
  \label{fig:beamfreq}
\end{figure}

Figure~\ref{fig:beamfreq} shows that the product $\theta_\mathrm{b}
\nu$ varies by $< \pm 3$\% across the band for both the vertical and
horizontal cuts through the beam.  The best L-band approximations with fixed
$\theta_\mathrm{b} \nu$ are:
\begin{equation}
  \begin{aligned}
  \theta_\mathrm{b} &= 89\,\farcm5  \left(\frac{\nu}{\mathrm{GHz}}\right)^{-1}
    \qquad && \mathrm{(Vertical)} \\
  \theta_\mathrm{b} &= 86\,\farcm2 \left(\frac{\nu}{\mathrm{GHz}}\right)^{-1}
    \qquad && \mathrm{(Horizontal)}.
  \end{aligned}
  \label{eqn:beamfreq}
\end{equation}

\section{DEEP2 Field Selection} \label{sec:selection}

The finite dynamic range of MeerKAT limits the minimum rms fluctuation
that can be achieved because every deep L-band image
contains thousands of sources distributed over the primary beam
area. The dynamic range of such images is defined by the ratio of the
effective source flux density $S_\mathrm{eff}$
to the rms fluctuation $\sigma$ in regions devoid of sources:
\begin{equation}
\mathrm{DR} \equiv \frac{S_\mathrm{eff}}{\sigma}~,
\label{eqn:dynrange}
\end{equation}
where $S_\mathrm{eff}$ is the quadratic sum of attenuated
flux densities $S_\mathrm{a}$ of all sources in the primary beam \citep{condon09}
\begin{equation}
S_\mathrm{eff} = \left(\sum_i S_{\mathrm{a,}i}^2 \right)^{\frac{1}{2}}~.
\label{eqn:seff}
\end{equation}
The attenuated flux density of a source with true flux density $S$
offset from the pointing center by an angle $\rho$ is
\begin{equation}
S_\mathrm{a} \equiv a_\mathrm{b}(\rho) S~,
\end{equation}
where
$a_\mathrm{b}(\rho)$ is the primary beam attenuation pattern approximated by a
circular Gaussian.
The quadratic sum over $S_\mathrm{a}$ in Equation~\ref{eqn:seff} is
primarily determined by the strongest sources in the field of view, so
the best area for a deep field has the fewest and faintest bright
sources in the primary beam.

The attenuated flux density $S_\mathrm{a}$ of a source varies with telescope
pointing errors and receiver gain fluctuations, both of which contribute to
$\sigma$.  Pointing errors contribute a flux-density error
\begin{equation}
\Delta S_\mathrm{p} (\rho) \approx \epsilon(\rho) S_\mathrm{a}~,
\end{equation}
where
\begin{equation}
\epsilon(\rho) \approx \frac{8\ln(2)\rho\sigma_\mathrm{p}}{\theta_\mathrm{b}^2}
\end{equation}
is the fractional attenuation change at an angle $\rho$ from the pointing center
caused by an rms pointing error $\sigma_\mathrm{p}$ \citep{condon09}.
Receiver gain fluctuations cause a flux-density change
\begin{equation}
\Delta S_\mathrm{g}(\rho) \approx \sigma_\mathrm{g} S_\mathrm{a}~,
\end{equation}
where $\sigma_\mathrm{g}$ is the rms receiver gain error.

Each source in an image contributes an error flux density which is the
quadratic sum of these flux-density fluctuations:
\begin{equation}
{\Delta}S = \sqrt{{\Delta}S^{2}_\mathrm{p} + {\Delta}S^{2}_\mathrm{g}}.
\label{eqn:deltas}
\end{equation}
Image artifacts produced by individual sources are independent, so we
define an image ``demerit'' score $d$ equal to the quadratic sum of the
independent source contributions to $\Delta S$ in the primary beam:
\begin{equation}
d \equiv \left(\sum_i {\Delta}S^{2}_i\right)^\frac{1}{2}~.
\label{eqn:demerit}
\end{equation}
To locate the best deep fields observable by MeerKAT, we searched the
Sydney University Molonglo Sky Survey \citep[SUMSS;][]{mauch03}
catalog south of $\delta = -35\degr$ and the NRAO VLA Sky Survey
\citep[NVSS;][]{condon98} catalog from $\delta \geq -35\degr$ to
$\delta = +10\degr$ over a fine grid of $\approx 6\times10^7$ potential
pointings separated by $1\farcm05$ in the $18185\,\deg^2$ area defined
by $\delta<10\degr, |b|>10\degr$.  At each grid position we computed
$d$ from flux densities shifted to 1.28 GHz assuming $S\propto\nu^{-0.7}$
and with parameters conservatively appropriate to MeerKAT:
$\theta_\mathrm{b} = 68\arcmin$, $\sigma_\mathrm{p} = 30\arcsec$, and
$\sigma_\mathrm{g} = 0.01$ \citep{jonas16} out to the radius $\rho =
3^\circ$ extending beyond the second sidelobe of the MeerKAT primary
beam.

\begin{table}
  \centering
  \caption{Five minimum-demerit positions with $d<1.4$\,mJy}
  \label{tab:demerit}
  \begin{tabular}{cccc}
    \hline
    $\alpha$ & $\delta$ & $d$ & \llap{$\Omega$} ($d<1.4$\,mJy\rlap{)} \\
    \multicolumn{2}{c}{(J2000)} & (mJy) & (deg$^2$) \\
    \hline
    $03^\mathrm{h}21^\mathrm{m}$ & $-18\degr53'$ & $1.32$ & $0.19$ \\
    $04^\mathrm{h}22^\mathrm{m}$ & $-80\degr15'$ & $1.35$ & $0.22$ \\
    $16^\mathrm{h}37^\mathrm{m}$ & $-70\degr46'$ & $1.34$ & $0.36$ \\
    $21^\mathrm{h}04^\mathrm{m}$ & $-54\degr25'$ & $1.36$ & $0.25$ \\
    $22^\mathrm{h}03^\mathrm{m}$ & $-35\degr43'$ & $1.34$ & $0.23$ \\
    \hline
  \end{tabular}
\end{table}

The five pointing centers with the smallest demerit flux densities $d$
in the southern hemisphere are listed in Table~\ref{tab:demerit},
along with their solid angles $\Omega$ in which $d <
1.4\,\mathrm{mJy}$. We chose the southernmost field at J2000 ${\alpha
  = 04^\mathrm{h}22^\mathrm{m}, \delta = -80\degr15\arcmin}$ to ensure
observations of the field could be easily scheduled at most times of
the day during the early commissioning and engineering phase of the
telescope. Also, at ecliptic latitude $\beta \approx -75^\circ$, the
DEEP2 field is easily observed by orbiting telescopes and is minimally
affected by zodiacal dust.  We inspected a mosaic image made in 2017
with a 16 antenna MeerKAT subarray and covering a 2\,deg$^2$ region
surrounding our selected position, and we finally chose the field
centered at J2000 ${\alpha = 04^\mathrm{h}13^\mathrm{m}26\,\fs 4,\,
  \delta = -80\degr00\arcmin00\arcsec}$ (which we call DEEP2) in order
to move a few moderately bright extended sources farther from the
pointing center.  Our DEEP2 field has a demerit score $d=1.4$\,mJy,
only slightly higher than the 1.35\,mJy minimum at J2000 ${\alpha =
  04^\mathrm{h}22^\mathrm{m}, \delta = -80\degr15\arcmin}$.

\section{Observations and Imaging} \label{sec:obsim}

\subsection{DEEP2 Observations}

The DEEP2 field centered on J2000
$\alpha=04^\mathrm{h}\,13^\mathrm{m}\,26\,\fs4$,
$\delta=-80\degr\,00\arcmin\,00\arcsec$ was observed at L band in 12
separate sessions between 2018 April 27 and 2019 January 20 for a total of
155.2\,hours (Table~\ref{tab:obs_summary}). We
always required that at least 58 of the 64 antennas and at least 7 of
the 9 outer-ring antennas (those providing the longest baselines) be
available. This ensured sufficient long- and intermediate-baseline
coverage to produce a fairly clean ``dirty beam'' point spread
function (PSF) with $\theta \lesssim 8''$ FWHM resolution.  We preferentially
observed during the night, though sessions with longer duration and
other scheduling constraints meant that $\sim30$\% of the observations
occurred in daytime. The $-80\degr$ declination of the DEEP2 field
ensures that it is never $< 55\degr$ from the Sun.

\begin{table}
  \centering
  \caption{DEEP2 observation summary}
  \label{tab:obs_summary}
  \begin{tabular}{lcccc}
    \hline
    \multicolumn{1}{c}{Date} & Start Time & $\tau_\mathrm{Total}$ & $\tau_\mathrm{Target}$ & N$_\mathrm{Ants}$ \\
          & UTC &  \multicolumn{2}{c}{(h)} & \\
    \hline
    2018 Apr 27 & 07:11 & 11.0 & \hphantom{1}8.4  & 61 \\
    2018 Jun 30 & 23:00 & 16.2 & 12.6 & 60 \\
    2018 Jul 7  & 21:39 & 17.2 & 13.4 & 61 \\
    2018 Jul 16 & 21:37 & \hphantom{1}8.0  & \hphantom{1}6.0  & 61 \\
    2018 Jul 24 & 21:07 & \hphantom{1}8.9  & \hphantom{1}6.9  & 59 \\
    2018 Jul 25 & 21:01 & \hphantom{1}9.0  & \hphantom{1}7.6  & 58 \\
    2018 Jul 27 & 21:01 & 16.1 & 14.0 & 61 \\
    2018 Jul 28 & 20:51 & 16.2 & 14.1 & 60 \\
    2018 Oct 8   & 21:33 & \hphantom{1}9.5  & \hphantom{1}8.5  & 59 \\
    2018 Nov 4   & 14:37 & 16.2 & 14.2 & 62 \\
    2019 Jan 19  & 09:31 & 16.1 & 13.9 & 63 \\
    2019 Jan 20  & 09:41 & 10.8 & \hphantom{1}9.2  & 63 \\
    \hline
    \multicolumn{1}{c}{Total} & & \llap{1}55.2 & \llap{1}28.8 &  \\
    \hline
  \end{tabular}
\end{table}

Scans on the DEEP2 target lasted 15\,minutes and were interleaved with scans
on the ${S(1284\mathrm{\,MHz}) \approx 6.1}$\,Jy phase and secondary
gain calibrator PKS\,J0252$-$7104 located $\approx10\degr$ from the DEEP2
field center.  Scans on PKS\,J0252$-$7104 were 2\,minutes long
before July 25, after which we shortened them to 1\,minute
because we found that we were getting sufficient SNRs on the
calibrator gain solutions. The primary flux-density and bandpass calibrator
PKS\,B1934$-$638 was observed for 10\,minutes at the start of each
observation and then subsequently every 3\,hours until it set.  Its
assumed flux densities from \citet{reynolds94} are listed as a function of
frequency in Table~\ref{tab:calsubbands}.

Table~\ref{tab:obs_summary} summarizes our observations of the DEEP2
field. From the total 155.2\,hours, calibration/slewing overheads took
17\% and left 128.8\,hours on the DEEP2 target.  Our
observations had an integration period of 8\,s except for the initial
April 27 observation that we averaged from 4\,s to 8\,s prior
to any calibration. Observations were all carried out in the $4096
\times 208.984$\,kHz channel L-band continuum mode.  The raw data
volume was typically $\approx 2\,\mathrm{TB}$ for a 12\,hour
observation.  Each dataset was calibrated separately prior to imaging.

The uncalibrated visibilities from our observations are publicly available and
were obtained from the SARAO archive at \url{https://archive.sarao.ac.za}.
Readers interested in obtaining these data can do so by searching
the archive for Proposal ID: SCI-20180426-TM-01.

\begin{table}
  \centering
  \caption{The 8 calibration subbands}
  \label{tab:calsubbands}
  \begin{tabular}{ccc}
    \hline
    Subband & $\nu$       & $S$(PKS\,B1934$-$638)\tablenotemark{a} \\
    number  &    (MHz)    &      (Jy)            \\
    \hline
    1       &  \hphantom{1}935.728    &     14.516 \\
    2       & 1035.204    &     14.921 \\
    3       & 1134.680    &     15.108 \\
    4       & 1234.157    &     15.129 \\
    5       & 1333.634    &     15.028 \\
    6       & 1433.110    &     14.835 \\
    7       & 1532.586    &     14.577 \\
    8       & 1632.063    &     14.428 \\
    \hline
  \end{tabular}
  \tablenotetext{a}{Flux densities from \citet{reynolds94}}
\end{table}

\begin{figure}
  \includegraphics[width=\columnwidth]{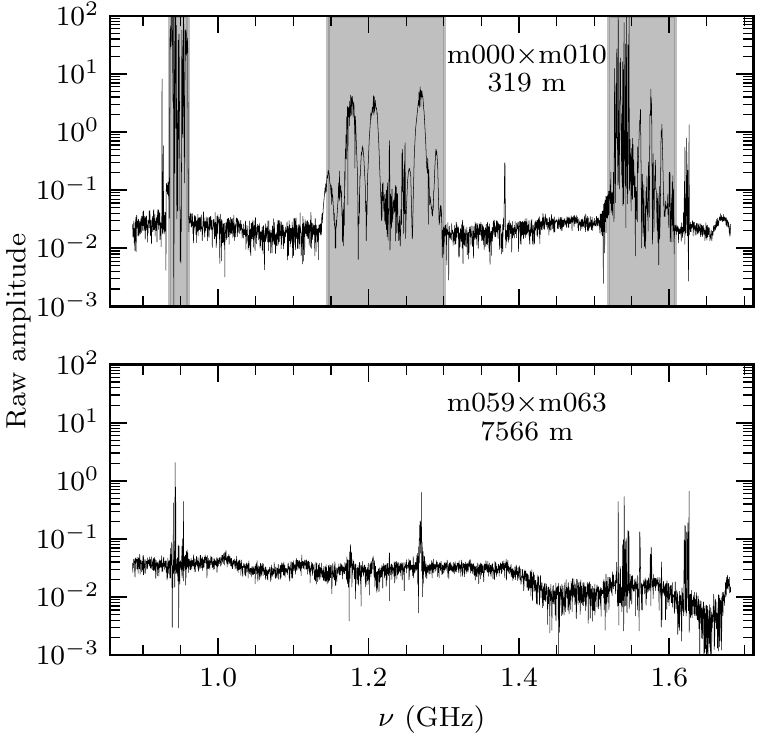}
  \caption{The average amplitude of cross-hand
           polarization visibilities (Horizontal $\times$ Vertical)
           from a 10\,minute scan on PKS\,B1934$-$638
           on two baselines. We chose to plot data from a cross-hand
           polarization as this is more sensitive to polarized RFI signals.
           A short baseline
           (m000$\times$m010; 319\,meters) is plotted in the upper panel and a
           long one (m059$\times$m063; 7566\,meters) is shown in the lower
           panel. The gray shaded areas in the upper panel show the regions
           masked for all times on baselines shorter than 1000\,meters.}
  \label{fig:mkatband}
\end{figure}

\subsection{Editing and Calibration}

The full MeerKAT L band covers the frequency range 856--1712\,MHz with
4096 spectral channels. We trimmed 144 channels each from the lower
and upper ends of the full band because the receiver response is too
weak at the band edges to be useful. The remaining frequency range
of our data is 886--1682\,MHz. This band is contaminated by various
sources of strong radio frequency interference (RFI), the broadest of
which are difficult to detect automatically. We therefore developed an
empirical mask that flags at all times those channels most
contaminated by RFI. Figure~\ref{fig:mkatband} shows an example of the
raw MeerKAT bandpass on two baselines, one short (319\,meters) and
one long (7566\,meters). The long baselines are typically not as badly
affected by RFI as the short ones, so we chose to apply the mask only
to baselines shorter than 1000\,m. The mask rejects 34\% of the
trimmed L band, or 17\% of the full dataset when applied only to the
subset of short baselines.

The unmasked data were searched in time and frequency for deviations
and flagged. Our flagging method is similar to the SumThreshold
technique described by ~\citet{offringa10}.  A smooth background was
fitted to the data by convolving them with a Gaussian whose widths in
frequency and time are larger than expected RFI spike widths and are
smaller than any variations in the bandpass or changes in amplitude
with time.  The smoothed background was then subtracted from the data
and outliers in the residuals were detected in increasing averaged
widths in both time and frequency.

After the data were masked and initially edited, we used the Obit
package \citep{cotton08} for further editing and calibration.  Prior
to calibration the raw data were smoothed with a Hanning filter to
prevent Gibbs ringing. This filter combines adjacent channels with
weights $(1/4,1/2,1/4)$ and effectively doubles the channel width to
$2\times208.984 \mathrm{~kHz} = 417.968$\,kHz.  It makes neighboring
frequency channels degenerate, so we excised every second channel from
the smoothed data prior to calibration.

Our calibration consisted of the following steps:
\begin{enumerate}
  \item The variations in group delays were calculated from the
    observations of PKS\,B1934$-$638 and PKS\,J0252$-$7104 and were
    interpolated to all of the data.

  \item A bandpass calibration was determined to remove residual
    variations in gain and phase as a function of frequency.  This was
    based on a CLEAN component model within $1\degr$ of
    PKS\,B1934$-$638. The average correction from all scans on
    PKS\,B1934$-$638 was applied to the data.

  \item The observations of PKS\,J0252$-$7104 were used to correct the
    phases and amplitudes on the target for each calibration subband
    as a function of time. The amplitudes of PKS\,B1934$-$638 were
    derived from the model of \citet{reynolds94} in each subband (see
    column 3 of Table~\ref{tab:calsubbands}) and used to determine the
    amplitude spectrum of PKS\,J0252$-$7104. Amplitude and phase
    corrections were then interpolated in time and applied to the
    entire dataset.

  \item Some residual errors that were not detected during earlier
    editing were found by searching for gains with amplitudes that are
    discrepant by more than $20\sigma$ and flagging them.

  \item The fully calibrated data were edited once more. At this stage
    the few visibilities with extremely discrepant Stokes I amplitudes
    ($>200$\,Jy) and outliers from a running median in time and
    frequency were flagged.

  \item The previous steps were repeated after resetting the
    calibration while retaining the flags on the calibrated data.
\end{enumerate}

After the editing steps listed above, $\sim35$\% of the data
were flagged. Prior to imaging, the calibrated visibilities were
spectrally averaged again, to ${2\times417.968 \mathrm{\,kHz} = 835.936
\mathrm{\,kHz}}$. The calibrated, flagged, and averaged data volume for
the target was $\approx 200$\,GB in a typical 12\,h observation.

\subsection{Self Calibration and Final Editing}

The DEEP2 data were collected during 12 observing sessions spread over
9 months (Table~\ref{tab:obs_summary}) and were individually phase
referenced to PKS\,J0252$-$7104, which is $\sim 10\degr$ from the
DEEP2 field center at J2000 ${\alpha = 04^\mathrm{h}\,13^\mathrm{m}
\, 26\,\fs4}$, ${\delta = -80^\circ \, 00\arcmin \,00 \arcsec}$.  The
resulting phases were corrected to the direction and times of the
target observations and astrometrically aligned with each other before
imaging.  The data from one day (2018 July 27)  were imaged using
two iterations of phase-only self-calibration with a 30\,s averaging
time.  The resulting model of the DEEP2 field was used
as the initial model to start the phase self-calibration of all data sets.
Starting the self-calibrations of other days from one model ensures that
images from all days are astrometrically aligned.
We avoided amplitude self-calibration for two reasons:
(1) amplitude self-calibration does not work well in a field containing
many faint sources and no dominant point source and (2)
the external gain calibration based on observations of PKS B1934$-$638
worked very well.  We measured the
flux density of the $S \approx 12 \mathrm{~mJy}$ point source
at J2000 $\alpha = 04^\mathrm{h}\,15^\mathrm{m}\,08\fs21$, $\delta
= -79^\circ\,59\arcmin\, 41\farcs0$ on the 12 daily DEEP2 images
(Table~\ref{tab:obs_summary}); its rms variation is only 2\%
over the full 9 month observation period.

The data from each session were
then imaged and deconvolved without further self-calibration to ensure
good data quality.  As a final editing step, the models derived from
these preliminary images were Fourier transformed and subtracted from
the calibrated $(u,v$) data.  Residual amplitudes $>0.5\mathrm{\,Jy}$
in the difference data were flagged in the calibrated session data.
Next the data were time averaged in a baseline--dependent
fashion using Obit/UVBlAvg with the constraints of $<1$\% amplitude
loss within $1\fdg5$ of the field center and averaging time $<30
\mathrm{~s}$.  Finally, the averaged data sets were
concatenated by Obit/UVAppend into a single data set for imaging.

\subsection{Imaging}

The concatenated data set was imaged using the Obit task MFImage
\citep{cotton18} without further self-calibration.  MFImage used small
planar facets to cover the wide field of view and made separate images
in the 14 imaging subbands (Table~\ref{tab:subbands}) having
fractional bandwidths $\Delta \nu / \nu < 0.05$ small enough to
accommodate the frequency dependence of sky brightness and primary
beam attenuation.  Dirty/residual images were formed in each subband,
but a weighted average image was used to drive the CLEAN process.
CLEAN components included the pixel flux densities from each subband,
and the subtraction during the major cycles used a spectral index
fitted to each component to interpolate between the subband center
frequencies listed in Table~\ref{tab:subbands}.  The joint
deconvolution of the subband images requires that the width of the
dirty PSF be nearly independent of frequency.  This was accomplished
by a frequency-dependent $(u,v)$ taper that downweighted the longer
baselines at the higher frequencies.

The facet images were re-projected during gridding to form a coherent
grid of pixels on the plane tangent to the celestial sphere at the
pointing center.  This allows a joint CLEAN of many facets in a given
major cycle.

The MeerKAT antenna array is centrally concentrated, so a relatively
strong Robust weighting ($\mathrm{ROBUST} = -1.3$ in AIPS/Obit usage)
was used to downweight the shortest baselines to give a
$\theta = 7\,\farcs 6$ FWHM synthesized beamwidth.  The DEEP2 field
was completely imaged out to a radius of $1\fdg5$ and facets out to
2$^\circ$ were added as needed to cover outlying strong sources
selected from the SUMSS \citep{mauch03} catalog.  The source density
in our DEEP2 image is so high that no CLEAN windowing was used.

CLEAN used a loop gain of 0.15 and found 250,000 point components
stronger than $7\,\mu\mathrm{Jy}$ for a total CLEAN flux density $S
= 1.301 \mathrm{\,Jy}$.  Prior to restoring the CLEAN components with a
circular Gaussian of FWHM $\theta = 7\,\farcs6$, the residuals
in each subband and facet were convolved with a Gaussian with widths
calculated to give a dirty PSF having approximately the same FWHM as
the restoring beam.
First the CLEAN components appearing in each facet of each subband
image were restored and then the subband facets were collected into a
single subband plane.  The output of this imaging process is an image
cube containing the 14 CLEANed and restored subband images.

\subsection{Wideband Images} \label{sec:wbim}

We took weighted averages of the subband images, all of which have
well-defined center frequencies $\nu_i$ and small fractional
bandwidths $\Delta \nu / \nu_i \lesssim 0.05$, to produce single-plane
wideband images using two different weighting schemes.  If the rms
noise in each subband image is $\sigma_i$, then  subband weights $w_i
\propto \sigma_i^{-2}$ minimize the wideband image noise variance
\begin{equation}
  \sigma_\mathrm{n}^2 = \sum_{i=1}^{14} w_i \sigma_i^2\, \Bigg/
  \sum_{i = 1}^{14} w_i~.
\label{eqn:minnoise}
\end{equation}
More generally, subband weights $w_i \propto (\nu_i^\alpha /
\sigma_i)^2$ maximize the wideband image SNR for sources with spectral
index $\alpha \equiv + d \ln (S)\,/\,d \ln(\nu)$:
\begin{equation}
  \mathrm{SNR}^2 \propto \sum_{i=1}^{14} w_i (\nu_i^\alpha / \sigma_i)^2\,
  \Bigg/ \sum_{i = 1}^{14} w_i~.
\label{eqn:maxsnr}
\end{equation}
Minimizing $\sigma_\mathrm{n}$ (Equation~\ref{eqn:minnoise}) is
equivalent to choosing $\alpha = 0$ in Equation~\ref{eqn:maxsnr}.  The
median spectral index of faint sources selected at frequencies $\nu
\sim 1.4 \mathrm{~GHz}$ is $\langle \alpha \rangle \approx -0.7$
\citep{condon84}, so this is the best choice of $\alpha$ for
maximizing the SNR.  The first three columns of
Table~\ref{tab:subbands} list the subband numbers, center frequencies
$\nu_i$ (MHz), and rms noise values $\sigma_i$
($\mu\mathrm{Jy~beam}^{-1}$).  Column~4 tabulates the weights $w_i$
that minimize $\sigma_\mathrm{n}^2$, and column~5 shows the
different weights $w_i$ that maximize the SNR in the wideband DEEP2
image.  Both sets of weights have been normalized to make $\sum_{i =
  1}^{14} w_i = 1$ for convenience.

\begin{table}
  \caption{DEEP2 imaging subband frequencies and weights}
  \label{tab:subbands}
  \begin{tabular}{r c c c c}
    \hline
    Subban{\rlap d} & $\nu_i$~~ & $\sigma_{\mathrm{n},i}$ & $w_i$ for & $w_i$ for \\
    numbe{\rlap r} & (MHz) & \llap{(}$\mu\mathrm{Jy~beam}^{-1}$\rlap{)} & min $\sigma_\mathrm{n}^2$ &
    max SNR \\
    \hline
 $i=1$ &    \hphantom{1}908.04 &  4.22 &  0.0225 &  0.0378 \\
     2 &    \hphantom{1}952.34 &  5.04 &  0.0158 &  0.0248 \\
     3 &    \hphantom{1}996.65 &  3.20 &  0.0393 &  0.0580 \\
     4 &   1043.46 &  2.88 &  0.0483 &  0.0669 \\
     5 &   1092.78 &  2.76 &  0.0526 &  0.0683 \\
     6 &   1144.61 &  2.58 &  0.0603 &  0.0733 \\
     7 &   1198.94 &  4.20 &  0.0227 &  0.0259 \\
     8 &   1255.79 &  3.98 &  0.0253 &  0.0271 \\
     9 &   1317.23 &  1.85 &  0.1171 &  0.1170 \\
    10 &   1381.18 &  1.64 &  0.1486 &  0.1389 \\
    11 &   1448.05 &  1.55 &  0.1672 &  0.1463 \\
    12 &   1519.94 &  1.87 &  0.1147 &  0.0938 \\
    13 &   1593.92 &  2.89 &  0.0481 &  0.0368 \\
    14 &   1656.20 &  1.85 &  0.1173 &  0.0850 \\
\hline
  \end{tabular}
  \end{table}

If a source near the pointing center has flux density
$S \propto \nu^\alpha$, its
flux density in the weighted wideband image will be
\begin{equation}
  S \propto \sum_{i = 1}^{14} w_i \nu_i^\alpha \Bigg/ \sum_{i = 1}^{14} w_i~.
\end{equation}
 We define the ``effective'' frequency $\nu_\mathrm{e}$ of a weighted
 wideband image as the frequency at which the image flux
 density equals the source flux density $\nu_\mathrm{e}^\alpha$:
\begin{equation}
  \nu_\mathrm{e} =
  \Biggl( \sum_{i = 1}^{14} w_i \nu_i^\alpha \Bigg/
  \sum_{i = 1}^{14} w_i \Biggr)^{1/\alpha}~.
\end{equation}
Thus different weighting schemes yield slightly different effective frequencies
for the wideband images.  The effective frequencies of our DEEP2
images are ${\nu_\mathrm{e} \approx 1329 \mathrm{~MHz}}$ for minimum
$\sigma_\mathrm{n}^2$ and $\nu_\mathrm{e} \approx 1278 \mathrm{~MHz}$
for maximum SNR weighting.

\begin{figure}
  \includegraphics[trim={4.6cm 9.7cm 5.1cm 9.7cm}, clip,
    width=\linewidth]{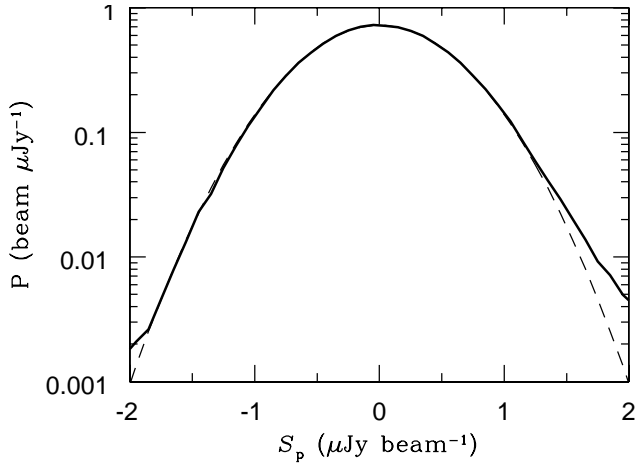}
  \caption{The distribution of peak flux densities of the SNR-weighted wideband
    image in five apparently source-free regions covering 14,373 synthesized
    beam solid angles $\sim 1 \fdg 6$ from the pointing center is shown by
    the thick curve.  It is well matched by the Gaussian with rms
    ${\sigma_\mathrm{n} = 0.55 \, \mu\mathrm{Jy\,beam}^{-1}}$ shown plotted as a broken curve.
    The Gaussian is a parabola on this semilogarithmic plot.}
  \label{fig:rmsnoise}
\end{figure}

The rms noise in the SNR-weighted DEEP2 image was estimated in five
widely separated and apparently source-free regions covering $14,373$
CLEAN beam solid angles at offsets $\rho \sim 1\fdg 6$ from the
pointing center, where the primary attenuation is $a_\mathrm{b} < 0.01$.  The
distribution of their peak flux densities $S_\mathrm{p}$ is nearly
Gaussian (Figure~\ref{fig:rmsnoise}) with rms $\sigma_\mathrm{n} = 0.55 \pm
0.01 \,\mu\mathrm{Jy\,beam}^{-1}$.  The noise in a synthesis image not
corrected for primary beam attenuation should be uniform across the
whole image.

\begin{figure}
  \includegraphics[trim={4.0cm 13.0cm 4cm 7.0cm}, clip,
    width=\columnwidth]{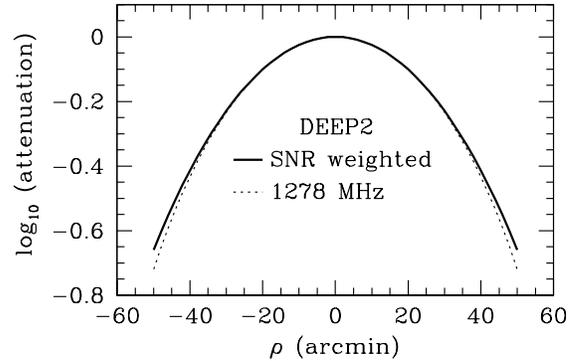}
  \caption{The circularized effective primary attenuation pattern for
    the wideband DEEP2 image made with subband weights that maximize
    the SNR for sources with $\alpha = -0.7$ (solid curve) has
    slightly broader wings than the narrow-band  attenuation pattern
    at $\nu_\mathrm{e} \approx 1278
    \mathrm{~MHz}$ (dotted curve).}
	\label{fig:atten}
\end{figure}

In each narrow subband the primary beam attenuation pattern
$a_{\mathrm{b},i}(\rho)$ is well defined and was measured accurately in the
horizontal and vertical planes (Section~\ref{subsubsec:primarybeam})
of the alt-az mounted MeerKAT dishes.  The primary beam is slightly
elliptical and rotates with parallactic angle on the sky.  For the
long DEEP2 tracks we approximated the ellipse by a circle whose
diameter is the geometric mean of the horizontal and vertical
diameters.  The primary beamwidth is inversely proportional to
frequency, so the effective primary pattern $a_\mathrm{b}(\rho)$ of the weighted
wideband image must be calculated from
\begin{equation}
  a_\mathrm{b}(\rho) = \sum_{i = 1}^{14} w_i a_{\mathrm{b},i}(\rho) \nu_i^\alpha \Bigg/
  \sum_{i=1}^{14} w_i ~.
\end{equation}
The circularized attenuation pattern for the wideband
DEEP2 image weighted for maximum SNR is shown in Figure~\ref{fig:atten}. Its
FWHM is $\theta_\mathrm{b} \approx 68\arcmin$.
Numerically it can be approximated within 0.1\% for all $a_\mathrm{b} > 0.25$ by
the polynomial
\begin{eqnarray}
&  a_\mathrm{b}(\rho)  \approx   1.0 - 0.3514 x / 10^3 + 0.5600 x^2 / 10^7 - \qquad\qquad
 \nonumber \\
& \llap{0.}0474 x^3 / 10^{10} +0.00078 x^4 / 10^{13} +0.00019 x^5 / 10^{16}\llap{,}\hphantom{00}
\label{eqn:priatten}
\end{eqnarray}
where $x \equiv [\rho\mathrm{(arcmin)}\nu_\mathrm{e}
  \mathrm{(GHz)}]^2$ and $\nu_\mathrm{e} \approx 1.278 \mathrm{~GHz}$.
This primary beam attenuation can be used by the AIPS task PBCOR via
the parameter PBPARM = 0.250, 1.0, $-0.3514$, 0.5600, $-0.0474$,
0.00078, 0.00019.  The wideband attenuation pattern is close to the
narrowband attenuation pattern at $\nu = \nu_\mathrm{e} \approx 1278
\mathrm{~MHz}$ shown by the dotted curve in Figure~\ref{fig:atten},
but it has slightly broader wings.

\begin{figure}
  \includegraphics[trim={0.0cm 0.0cm 0.0cm 0.0cm}, clip,
    width=\columnwidth]{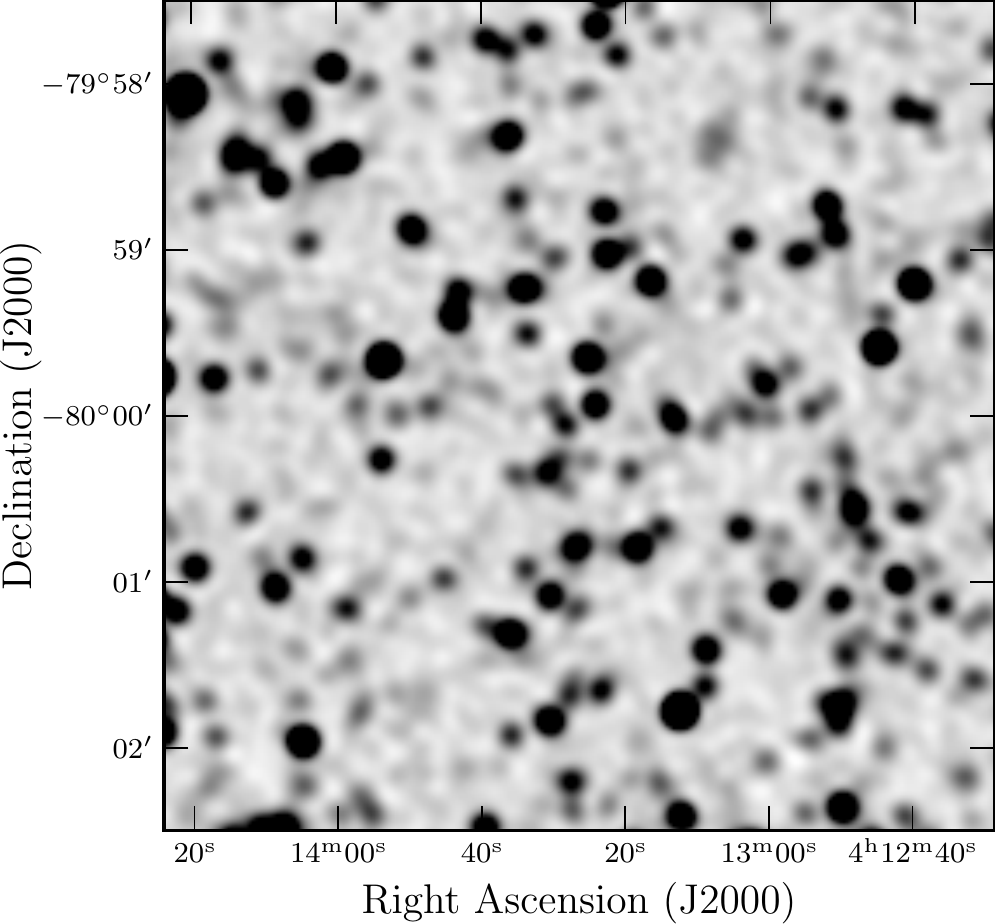}
 \caption{The central $5\arcmin \times 5\arcmin$ of the wideband
    DEEP2 image made with subband weights that maximize the SNR for
    sources with $\alpha = -0.7$ is covered by sources unresolved by
    the $\theta = 7\,\farcs6$ beam and brighter than the rms
    noise $\sigma_\mathrm{n} \approx 0.55 ~\mu\mathrm{Jy~beam}^{-1}$.
    The $-1.4~\mu\mathrm{Jy~beam}^{-1}$ ``bowl'' (as described in
    Section~\ref{sec:wbim}) has been removed from this image.}
	\label{fig:postage}
\end{figure}

\begin{figure*}
  \includegraphics[width=\textwidth]{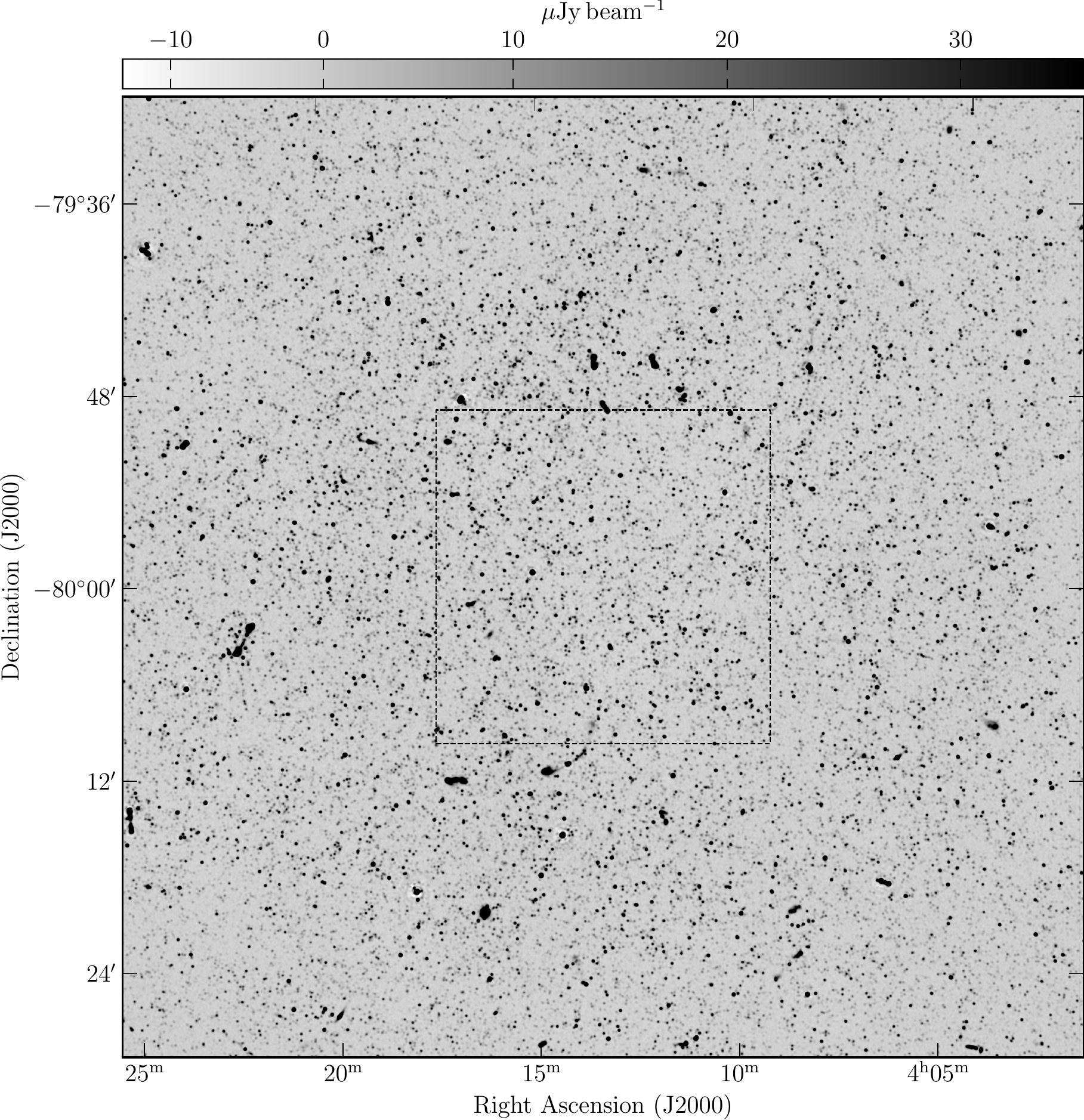}
  \caption{The central $1\degr \times 1\degr$ of the wideband DEEP2
    sky image made with subband weights that maximize the SNR for
    sources with $\alpha = -0.7$. The $-1.4~\mu\mathrm{Jy~beam}^{-1}$
    bowl described in Section~\ref{sec:wbim} has been removed from
    this image, and it has also been corrected for the primary beam
    attenuation using Equation~\ref{eqn:priatten}. The grey scale is
    stretched by an exponent of 1.3 between $-15\,\mathrm{{\mu}Jy}$
    and $30\,\mathrm{{\mu}Jy}$ as indicated by the bar at the
    top. The dashed square in the centre of the image bounds the
    $1250\arcsec\times1250\arcsec$ region whose $P(D)$ distribution we
    used to calculate the power-law source count described in
    Section~\ref{sec:powerlawpd}.}
  \label{fig:deep2image}
\end{figure*}

The wideband DEEP2 images are so sensitive (${\sigma_\mathrm{n} \approx
0.55 ~\mu\mathrm{Jy~beam}^{-1}}$) that they are densely covered by
sources with flux densities $S \gg \sigma_\mathrm{n}$
(Figures~\ref{fig:postage} and \ref{fig:deep2image}).  The total flux
density in a dirty interferometer image is always zero because there
are no zero-spacing $(u,v)$ data, only sinusoidal fringes with zero
means.  The dirty image of a single strong point source contains a
negative ``bowl'' whose angular size is inversely proportional to the
smallest $(u^2 + v^2)^{1/2}$ sampled.  For our DEEP2 data, the bowl
surrounding each source is much wider than the source spacing but much
narrower than the primary attenuation pattern $a_\mathrm{b}(\rho)$
(Figure~\ref{fig:atten}).  Faint sources have a fairly uniform random
distribution on the sky, and their attenuated brightness distribution
in the DEEP2 images is multiplied by the primary attenuation
pattern. Consequently each dirty DEEP2 image has a negative bowl whose
size and shape closely matches the primary beam and whose depth
exceeds $\sigma_\mathrm{n}$.  Partial CLEANing reduces the depth of
the bowl but does not completely eliminate it.

We estimated the central depth of the bowl of our SNR-optimized DEEP2
image using the mode of the brightness distribution in the $6' \times 6'$
square at the center the wideband CLEAN image, where the
primary attenuation is confined to the narrow range $0.98 < a_\mathrm{b}< 1.00$
(Figure~\ref{fig:atten}); it is $ -1.4\pm 0.1~\mu\mathrm{Jy~beam}^{-1}$.
To fill in the bowl and flatten the image
baseline level, we added the weighted wideband primary attenuation
pattern $a_\mathrm{b}(\rho)$ (Equation~\ref{eqn:priatten}) multiplied
by $1.4 \times 10^{-6}$ to the wideband image, whose brightness units
are Jy~beam$^{-1}$.

The central square cutout $4640 \mathrm{\,pixels} \times 1\farcs25
\mathrm{\,pixel}^{-1} = 5800\arcsec$ on a side from the SNR-weighted
1.28\,GHz wideband DEEP2 image with the bowl removed, but not
corrected for primary-beam attenuation, is available in FITS format at
\url{https://archive.sarao.ac.za}.

\section{Source Counts at 1.4~GHz} \label{sec:counts}

We used the DEEP2 confusion $P(D)$ distribution to make the best
power-law approximation to the sky density of sources fainter than
$10\,\mu\mathrm{Jy}$, and we counted individual DEEP2 sources
with $-5.0 < \log[S\mathrm{(Jy)}] < -2.6$.

\subsection{Deep Power-law P(D) counts}\label{sec:powerlawpd}

The differential number $n(S)$ of sources per unit flux density per
steradian is a statistical quantity that can be calculated directly
from the distribution of image brightnesses expressed in units of flux
density per beam solid angle.  The term ``confusion'' describes the
brightness fluctuations caused by radio sources, and an image is said
to be confusion limited if the confusion fluctuations are larger than
the rms noise fluctuations $\sigma_\mathrm{n}$.  The normalized probability
distribution of confusion brightness is traditionally called the
$P(D)$ distribution, where $D$ originally stood for the pen deflection
on a chart recording but now refers to image peak flux density
$S_\mathrm{p}$; that is, flux density per beam solid angle.

The noiseless $P(D)$ distribution for point sources can be derived
analytically only for power-law differential source counts $n(S) = k
S^{-\gamma}$, where $k$ is the overall source density parameter and
$\gamma$ is the power-law exponent \citep{condon74}.  Power-law
distributions are scale-free, so the shape of a power-law $P(D)$ distribution
depends only on $\gamma$.  The amplitudes and widths of power-law
$P(D)$ distributions obey the scaling relation
\begin{equation}
  P[(k \Omega_\mathrm{e})^{1/(\gamma-1)} D] =
  (k \Omega_e)^{-1/(\gamma-1)} P(D)~,
\label{eqn:pofdscaling}
\end{equation}
where
\begin{equation}
  \Omega_\mathrm{e} \equiv \int [a(\rho,\phi)]^{\gamma-1} \, d \Omega
\label{eqn:omegae}
\end{equation}
is the ``effective'' solid angle $\Omega_\mathrm{e}$ of a beam whose
polar attenuation pattern is $a(\rho, \phi)$.  For power-law counts,  the  $P(D)$
distribution depends on $\Omega_\mathrm{e}$ but not on the form of the beam attenuation pattern.  The
effective solid angle of an elliptical Gaussian beam with FWHM axes
$\theta_1$ and $\theta_2$ is
\begin{equation}
  \Omega_\mathrm{e} = \Biggl( \frac {\pi \theta_1\theta_2}{4 \ln 2} \Biggr)
  \Biggl( \frac {1} {\gamma - 1} \Biggr) =
  \frac {\Omega_\mathrm{b}}{\gamma - 1}~,
\label{eqn:omegaegauss}
\end{equation}
where $\Omega_\mathrm{b}$ is the Gaussian beam solid angle
\begin{equation}
  \Omega_\mathrm{b} \equiv \int a(\rho, \phi) \, d \Omega
  = \frac {\pi \theta_1 \theta_2}{4 \ln 2}~.
\label{eqn:omegab}
\end{equation}
A convenient normalization for displaying analytic $P(D)$ distributions
is $k \Omega_\mathrm{e} = \eta_1^{-1}$, where
\begin{equation}
  \eta_1^{-1} = \frac {2 \Gamma(\gamma/2) \Gamma[(\gamma+1)/2]
    \sin [\pi (\gamma -1) / 2]} { \pi^{\gamma + 1/2}}
\label{eqn:eta1}
\end{equation}
and $\Gamma(x)$ is the factorial function \citep{condon74}.

\begin{figure}
  \includegraphics[trim={2.0cm 6.8cm 5.0cm 3.5cm},
    clip, width=\columnwidth]{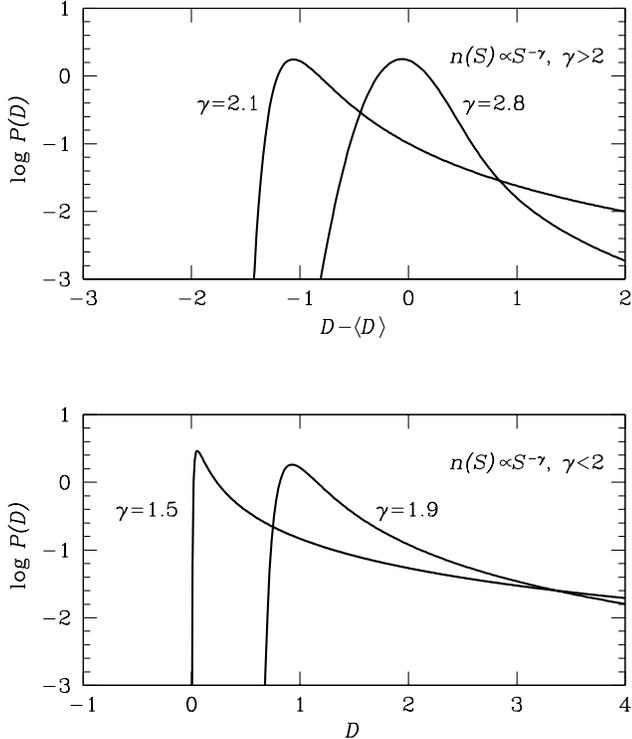}
  \caption{Noiseless $P(D)$ distributions for power-law
    source counts $n(S) \propto S^{-\gamma}$.  The top panel shows
    $\gamma > 2$ counts for which the sky brightness diverges (Olbers'
    paradox), so the deflections are shown relative to the mean deflection
    $\langle D \rangle$.  The bottom panel shows $P(D)$ distributions
    for $\gamma < 2$ relative to the absolute zero of the source sky.}
	\label{fig:pdcalc}
\end{figure}

The shape of the $P(D)$ distribution varies significantly with the count slope
$\gamma$. Four examples of $P(D)$ distributions with $k \Omega_\mathrm{e} =
\eta_1^{-1}$ are shown in Figure~\ref{fig:pdcalc}.  The
super-Euclidean slope $\gamma = 2.8$ is close to the actual
slope observed above $S \sim 1 \mathrm{~Jy}$ at 1.4~GHz,
so the $\gamma = 2.8$ curve in the upper
panel of Figure~\ref{fig:pdcalc} represents the troublesome confusion
that affected early radio surveys such as 2C \citep{shakeshaft55}.  In the limit $\gamma
\rightarrow 3-$, the $P(D)$ distribution is Gaussian and would appear
as a parabola in the semilogarithmic Figure~\ref{fig:pdcalc}.  The
$\gamma = 2.8$ curve is dominated by its nearly parabolic core and has
only a weak tail extending to the right.  The sky brightness
contributed by point sources diverges for all $\gamma \geq 2$ (Olbers'
paradox), so the calculated values of $D$ are relative to the mean
deflection $\langle D \rangle$.  The $\gamma = 2.1$ curve shows how
the $P(D)$ peak shifts to the left and the tail becomes more prominent
as $\gamma \rightarrow 2+$.

The lower panel of Figure~\ref{fig:pdcalc} shows sample $P(D)$
distributions when $\gamma < 2$ and $D$ represents the deflection
above the absolute zero of sky brightness contributed by radio
sources.  The $\gamma = 1.9$ curve is similar to the $\gamma = 2.1$
curve in form, and its peak $D$ is still significantly offset above zero by
contributions from faint sources so numerous that multiple sources are
blended together in each beam.  This is characteristic of confusion
seen at levels $10~\mu\mathrm{Jy} \lesssim S_\mathrm{1.4~GHz} \lesssim
0.1 \mathrm{~Jy}$.

At the lower value $\gamma \approx 1.5$ observed when $S_\mathrm{1.4~GHz}
\ll 10 \,\mu\mathrm{Jy}$, the nature of
confusion changes again.  There are so few sub-$\mu$Jy sources that
the $P(D)$ peak deflection approaches $D \rightarrow 0+$, indicating
that the extragalactic background has been largely resolved into discrete
sources. The long tail of the $P(D)$ distribution so completely
dominates the narrow core that the rms confusion $\sigma_\mathrm{c}$
is ill defined and should not be used to describe the amount of
confusion when $S_\mathrm{1.4~GHz} \ll 10\,\mu\mathrm{Jy}$.  Confusion
in the traditional sense ``melts away'' at the sub-$\mu$Jy levels
reached by DEEP2.  Instead of numerous even fainter sources tending to
boost the flux densities of faint sources, faint sources are more
likely to suffer obscuration by stronger sources.

\begin{figure}
  \includegraphics[trim={2.0cm 9.8cm 5.0cm 6.5cm},
    clip, width=\columnwidth]
    {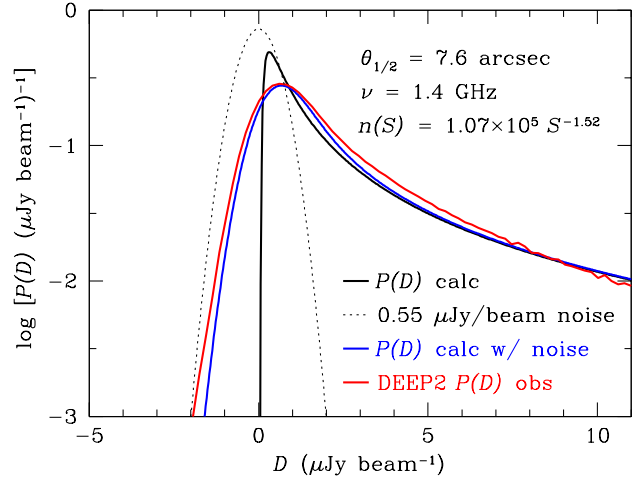}
  \caption{The solid black curve is the noiseless $P(D)$ distribution
    rescaled for a $\theta = 7\,\farcs6$ Gaussian beam and
    1.4~GHz source count
    $n(S) = kS^{-\gamma} = 1.07 \times 10^5 S^{-1.52}$.
    The dotted parabola is the normalized probability
    distribution of the $\sigma_\mathrm{n} = 0.55
    \,\mu\mathrm{Jy~beam}^{-1}$ Gaussian noise, and the blue curve is
    the convolution of the noiseless calculated $P(D)$ distribution
    with the noise.  The observed $P(D)$ distribution in the central
    $1250\arcsec \times 1250\arcsec$ ($\sim 2.4 \times 10^4
    \Omega_\mathrm{b}$) of DEEP2 is shown as the red curve.}
	\label{fig:pdall}
\end{figure}

\begin{figure}
  \centering
  \includegraphics[trim={2.2cm 8.9cm 6.5cm 4.3cm}, clip,
    width=\columnwidth]{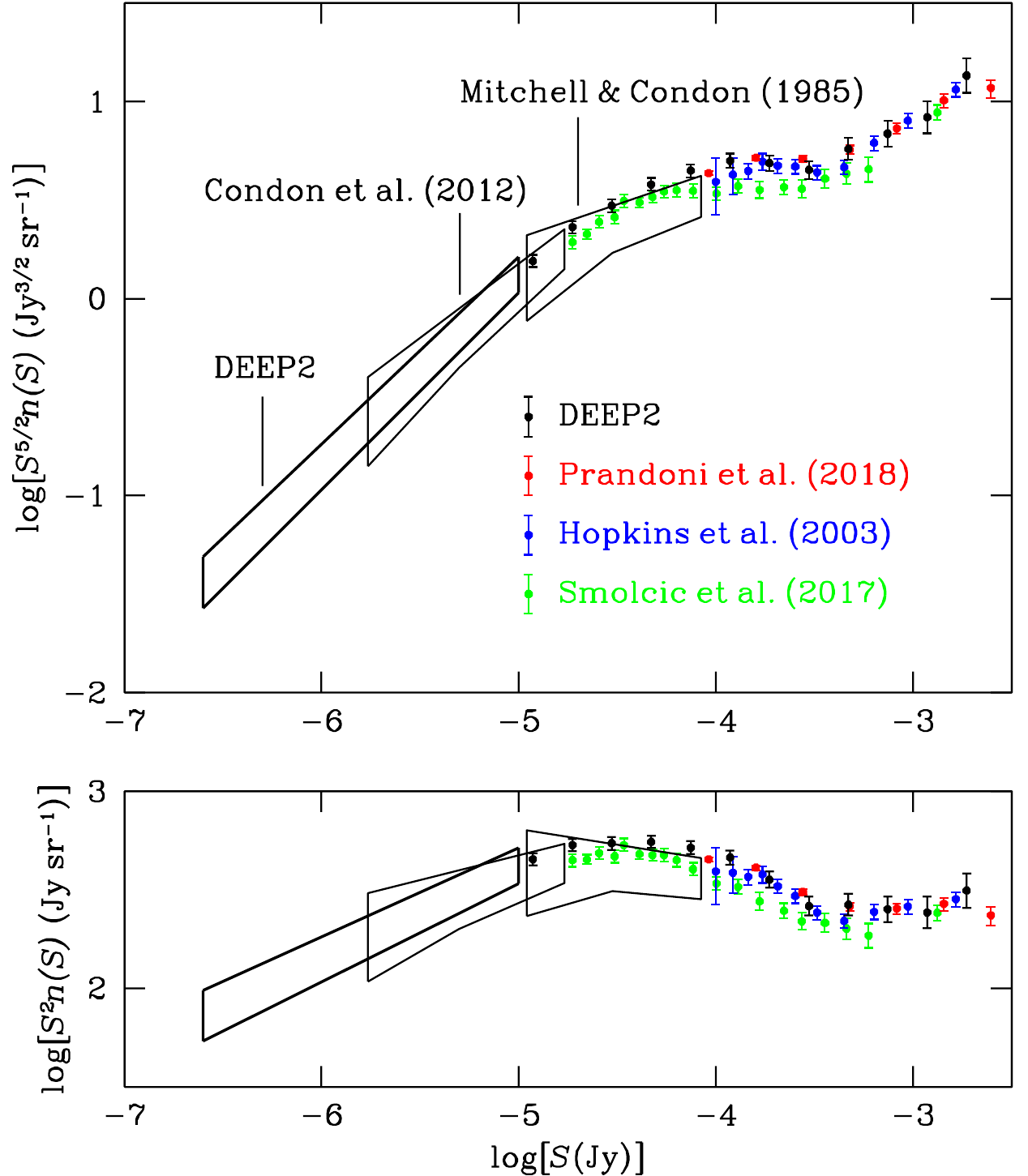}
  \caption{The 1.4 GHz differential source count has been plotted with
    the traditional static Euclidean normalization $S^{5/2}n(S)$ (top
    panel) and the brightness-weighted normalization $S^2 n(S)$
    (bottom panel). The black data points show the DEEP2 discrete
    source counts from Table~\ref{tab:deep2counts}, the red data points
    are from \citet{prandoni18}, the green data points are from
    \citet{smolcic17}, and the blue data points are from
    \citet{hopkins03}.
    The box covering $-5.0 < \log[S\mathrm{(Jy)}] < -4.1$
    bounds the \citet{mitchell85} 1.4~GHz $P(D)$ count and the
    box covering $-5.8 < \log[S\mathrm{(Jy)}] < -4.8$
    bounds the 3~GHz $P(D)$ count from
    \citet{condon12} converted to 1.4~GHz via spectral index $\alpha =
    -0.7$.  The heavy box
    spanning $-6.6 < \log[S\mathrm{(Jy)}] < -5$ indicates the best
    fit to the DEEP2 confusion $P(D)$ distribution discussed in
    Section~\ref{sec:powerlawpd}.
    }
	\label{fig:counts}
\end{figure}

The central $1250\arcsec \times 1250\arcsec$ square covering about
$2.4 \times 10^4$ restoring beam solid angles was extracted from the
DEEP2 sky image (Figure~\ref{fig:deep2image}) and rescaled to 1.4~GHz via
the spectral index $\alpha = -0.7$ typical of faint sources.
Its 1.4~GHz $P(D)$ distribution is shown by the red curve in
Figure~\ref{fig:pdall}. The best least-squares power-law fit with fixed rms noise
$\sigma_\mathrm{n} = 0.55 \pm 0.01 \, \mu\mathrm{Jy~beam}^{-1}$ but
free $k$ and $\gamma$ is $n(S) = 1.07 \times 10^5 S^{-1.52}
{\mathrm{Jy}^{-1}\,\mathrm{sr}^{-1}}$ between $S =
0.25\,\mu{\mathrm{Jy}}$ and $10\,\mu{\mathrm{Jy}}$.
Letting $\sigma_\mathrm{n}$ be a free parameter and varying
$\sigma_\mathrm{n}$ by $2 \Delta\sigma_\mathrm{n} =  \pm 0.02 \,\mu\mathrm{Jy~beam}^{-1}$
has a negligible effect on this fit.
The fit and its rms uncertainties (68\% confidence region) are bounded by
the thick box at the left in Figure~\ref{fig:counts}.  The actual source
count is not a perfect power law and the DEEP2 dirty beam is not perfectly
Gaussian, so getting more accurate source-counts will require
numerical simulations based on more realistic non-power-law source
counts and non-Gaussian dirty beams (Matthews et al., in prep).

Smoothly extrapolating our power-law 1.4\,GHz count below $S_0 \approx
0.25 \,\mu\mathrm{Jy}$ is reasonable because fainter sources are
evolved from the power-law region below the ``knee'' in the local RLF
\citep{condon19}, so simple evolutionary models \citep{condon84}
predict nearly power-law sub-$\mu$Jy source counts.  This
extrapolation yields an estimate of the Rayleigh-Jeans brightness
temperature $T_\mathrm{b}$ contributed by all fainter star-forming
galaxies:
\begin{equation}
  \Delta T_\mathrm{b}(< S_0) =
  \Biggl[ \frac {\ln(10) c^2} {2 k_\mathrm{B} \nu^2} \Biggr]
  \int_0^{S_0} S^2 n(S) d [\log{S}]~,
\end{equation}
where $k_\mathrm{B} \approx 1.38 \times 10^{-23} \mathrm{\,J\,K}^{-1}$
is the Boltzmann constant \citep{condon12}.
For $n(S) = 1.07 \times
10^5 S^{-1.52} \mathrm{~Jy}^{-1} \mathrm{~sr}^{-1}$ at 1.4\,GHz, the
background contributed by sources fainter than ${S_0 = 0.25
\,\mu\mathrm{Jy}}$ is $\Delta T_\mathrm{b} \approx 2.5
\mathrm{~mK}$, which is only $\approx 7$\% of the
$T_\mathrm{b} \approx 37 \mathrm{~mK}$ total background \citep{condon12}
contributed by star-forming galaxies and $<3$\% of the background
contributed by all extragalactic sources.

\subsection{Confusion and Obscuration Sensitivity Limits for
Individual Sources}

An image is said to be confusion limited if the errors caused by
confusion exceed the errors caused by Gaussian noise.  Both confusion
and noise set lower limits to the flux density $S$ of the faintest
individual source that can be reliably detected.
This section extends earlier calculations of the confusion limit
below $S \sim 10 \,\mu\mathrm{Jy}$, where $\gamma \ll 2$ and
obscuration dominates confusion.  This flux-density range
affects the confusion/obscuration corrections to DEEP2 direct source counts
(Section~\ref{sec:directcounts}) and directly impacts the design of future
arrays such as the SKA and ngVLA.

A common way to compare confusion and noise is to calculate their
variances $\sigma^2_\mathrm{c}$ and $\sigma^2_\mathrm{n}$.  This
calculation must be done carefully because the confusion variance
$\sigma_\mathrm{c}^2 = \int_0^\infty D^2 P(D) \,d D$ formally diverges
for all power-law source counts $n(S) = k S^{-\gamma}$ and is finite
only if the $P(D)$ distribution is truncated above some cutoff
deflection $D_\mathrm{c}$.  Then \citep{condon74}
\begin{equation}
  \sigma_\mathrm{c} = \Biggl( \frac {k \Omega_\mathrm{e}} {3 - \gamma}
  \Biggr)^{1/2} D_\mathrm{c}^{(3 - \gamma)/2}~,~~ 1 < \gamma < 3~.
\end{equation}
The cutoff should be proportional to $\sigma_c$: $D_\mathrm{c} = q\,
\sigma_\mathrm{c}$, where the constant $q \sim 5$ is the usual cutoff
signal-to-confusion ratio.  Then
\begin{equation}
  \sigma_\mathrm{c} = \Biggl( \frac {k \Omega_\mathrm{e}}
        {3 - \gamma} \Biggr)^{1/(\gamma-1)}
        q^{(3-\gamma)/(\gamma-1)}~.
\end{equation}
Note that the rms confusion $\sigma_\mathrm{c}$ still depends on
the choice of $q$.  Only as $\gamma \rightarrow 3-$ is $\sigma_\mathrm{c}$
nearly independent of $q$. When $\gamma = 2$, $\sigma_\mathrm{c} \propto q$.
In the sub-$\mu$Jy regime where $\gamma \sim 1.5$, $\sigma_\mathrm{c} \propto
q^3$ depends so sensitively on $q$ that the very concept of rms
confusion is nearly useless.

For imaging with a Gaussian beam, $\Omega_\mathrm{e} =
\Omega_\mathrm{b} / (\gamma - 1)$ and
\begin{equation}
  \sigma_\mathrm{c}^{\gamma -1} =
  \frac {k \Omega_\mathrm{b}\, q^{3-\gamma}} {(\gamma - 1)(3-\gamma)}~.
\label{eqn:sigc}
\end{equation}
Solving the cumulative source count
\begin{equation}
  N(>S) = \frac {k S^{1-\gamma}} {\gamma - 1}
\end{equation}
for $k = (\gamma-1) N(>S) S^{\gamma-1}$ and substituting the detection limit
$S \approx q \,\sigma_\mathrm{c}$ results in
\begin{equation}
  \Biggl( \frac {\sigma_\mathrm{c}} {S} \Biggr)^{\gamma-1} \approx
  \Biggl( \frac {q^{3-\gamma}} {3-\gamma} \Biggr) N(>S) \Omega_\mathrm{b} =
  \Biggl( \frac {q^{3-\gamma}} {3-\gamma} \Biggr) \beta^{-1}~,
\end{equation}
which can be solved for $\beta_\mathrm{min}$, the minimum number of beam solid
angles per reliably detectable source, in terms of the confusion
signal-to-noise ratio $q$:
\begin{equation}
  \beta_\mathrm{min}  \approx \frac {q^2}{3-\gamma}~.
\label{eqn:betamin}
 \end{equation}

\begin{figure}
  \includegraphics[trim={1.5cm 9.9cm 5.0cm 6.5cm},
    clip, width=\columnwidth]{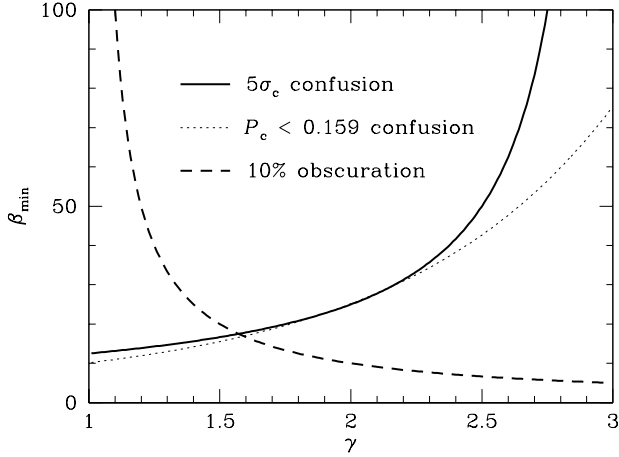}
  \caption{The solid curve is the minimum number $\beta_\mathrm{min}$
    of beam solid angles per source for a confusion SNR $q = 5$
    (Equation~\ref{eqn:betamin}), and the dotted curve shows an
    alternative $\beta_\mathrm{min}$ for which the probability of
    confusion is $P_\mathrm{c} < 0.159$ (the probability that Gaussian
    noise exceeds $+\sigma_\mathrm{n}$) for confusion $>S/5$
    (Equation~\ref{eqn:altbetamin}). The dashed curve is the minimum
    number of sources per beam solid angle at which 10\% of sources
    are obscured by stronger sources (Equation~\ref{eqn:betaminobsc}).}
	\label{fig:betamin}
\end{figure}

The number of beams per source at the $q = 5$ confusion limit is
shown as a function of $\gamma$ by the solid curve in
Figure~\ref{fig:betamin}.  In the limit $\gamma \rightarrow 3-$, the
$P(D)$ distribution is dominated by the very faintest sources and the
\emph{fluctuations} in sky brightness diverge, just as the total sky
brightness diverges as $\gamma \rightarrow 2-$ (Olbers' paradox).  The
$P(D)$ distribution becomes nearly Gaussian, the same as the noise
distribution, so sources can never be distinguished from noise and
$\beta \rightarrow \infty$.  For sources stronger than $S \sim 1
\mathrm{~Jy}$ at 1.4~GHz, a super-Euclidean differential count slope
$\gamma \approx 2.7$ implies $\beta_\mathrm{min} \approx 80$ beam
solid angles per reliable $q = 5$ source detection.  This very severe
requirement on $\beta_\mathrm{min}$ was not obvious when the first
extragalactic radio surveys were being made, so sources with smaller
$\beta$ were often cataloged as real and later found to be spurious.

An alternative approach to calculating the confusion
$\beta_\mathrm{min}$ uses the probability $P_\mathrm{c}$ that a source
of flux density $S$ will be confused by a weaker source of flux
density between $fS$ and $S$, where $f < 1$.  For a cumulative source
count $N(>S) \propto S^{1-\gamma}$ and a top-hat beam [$a(\theta,\phi)
  = 1$ over solid angle $\Omega_\mathrm{e}$ in the notation of
  Equation~\ref{eqn:omegae}],
\begin{equation}
  P_\mathrm{c} = \Omega_\mathrm{e} [ N(>fS) - N(>S)] =
  \Omega_\mathrm{e} [N(>S) (f^{1-\gamma} - 1)]~.
\end{equation}
A Gaussian beam with beam solid angle $\Omega_\mathrm{b}
= \Omega_\mathrm{e}  (\gamma - 1)$
yields exactly the same $P(D)$ distribution, so
\begin{equation}
  P_\mathrm{c} = N(>S) \Omega_\mathrm{b}
  \Biggl( \frac {f^{1-\gamma} - 1}{\gamma-1} \Biggr) =
    \frac {1} {\beta} \Biggl( \frac {f^{1-\gamma} - 1}{\gamma-1} \Biggr)
\end{equation}
and
\begin{equation}
  \beta_\mathrm{min} = \frac {1} {P_\mathrm{c}}
  \Biggl( \frac {f^{1-\gamma} - 1}{\gamma-1} \Biggr)
\label{eqn:altbetamin}
\end{equation}
is the minimum number of Gaussian beams per source consistent with
this confusion requirement.  A reasonable choice for $f$ would be $f
\approx 0.2$, so the confusion is at least $S/5$.  A reasonable choice
for the probability of confusion at this level is $P_\mathrm{c}
\approx 0.159$, the probability that Gaussian noise exceeds $+
\sigma_\mathrm{n}$.  The dotted curve in Figure~\ref{fig:betamin}
shows the resulting $\beta_\mathrm{min}$ as a function of $\gamma$.

In the broad flux-density range $10 ~\mu\mathrm{Jy} < S < 0.1
\mathrm{~Jy}$ covered by most 1.4~GHz surveys, $\gamma \sim 2$ and
${\beta_\mathrm{min} \sim 25}$.  Below $S \sim 10~\mu\mathrm{Jy}$,
the differential count slope falls again, to $\gamma \lesssim
1.5$. While $\beta_\mathrm{min}$ for avoiding \emph{confusion} by {\it
  fainter} sources continues to fall, the probability of
\emph{obscuration} by \emph{stronger} sources grows.  For a top-hat
beam  the probability that a source
of flux density $S$ will be obscured by a stronger source is
$P_\mathrm{o} = N(>S) \Omega_\mathrm{e}$.  For power-law source
counts, the $P(D)$ distribution is independent of beam shape and
depends only on $\Omega_{e}$, so for a Gaussian beam with beam solid
angle $\Omega_\mathrm{b} = (\gamma - 1) \Omega_\mathrm{e}$
(Equation~\ref{eqn:omegaegauss}),
\begin{equation}
  P_\mathrm{o} = N(>S) \,\Omega_\mathrm{e}
  = \frac {N(>S) \,\Omega_\mathrm{b}} {\gamma - 1}
\end{equation}
and the minimum number of beam solid angles per source to minimize
obscuration is
\begin{equation}
  \beta_\mathrm{min} \approx [P_\mathrm{o} (\gamma - 1)]^{-1}~.
\label{eqn:betaminobsc}
\end{equation}
The dashed curve in Figure~\ref{fig:betamin} shows the number
$\beta_\mathrm{min}$ of beam solid angles per source corresponding to
$P_\mathrm{o} = 0.1$.

\begin{figure}
  \includegraphics[trim={2cm 9.8cm 3.50cm 7.5cm},
    clip, width=\columnwidth]{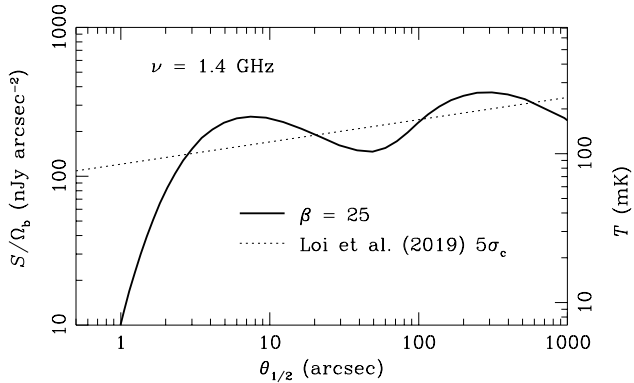}
  \caption{The solid curve shows the individual-source detection limit
    $S$ divided by the beam solid angle $\Omega_\mathrm{b}$ in units
    of nJy~arcsec$^{-2}$ (left ordinate) and Rayleigh-Jeans brightness
    temperature $\sigma_\mathrm{T}$ in units of mK (right ordinate) at
    1.4~GHz at which $\beta = 25$.  The dotted line is
    $5\sigma_\mathrm{c}$ calculated for the \citet{loi19} 1.4~GHz SKA
    sky simulation.}
\label{fig:sigconf2}
\end{figure}

Choosing $\beta_\mathrm{min} \approx 25$ is a good rule-of-thumb for
detecting individual sources valid in the flux-density range
below $S_\mathrm{1.4~GHz}
\sim 0.1 \mathrm{~Jy}$, where $1.4 \lesssim \gamma \lesssim 2$.
The solid curve in Figure~\ref{fig:sigconf2} shows the 1.4~GHz flux
density $S$ at which $\beta = 25$.  For
example, the 1.4~GHz EMU survey \citep{norris11} has $\theta =
10\arcsec$ FWHM resolution, rms confusion $S / \Omega_\mathrm{b} \approx
240 \mathrm{~nJy~arcsec}^{-2}$, and beam solid angle $\Omega_\mathrm{b} =
\pi \theta^2 / (4 \ln 2) \approx 113 \mathrm{~arcsec}^2$.
The weakest reliably detectable sources at that resolution have flux
densities $S \approx 27 ~\mu\mathrm{Jy}$.

The sharp decline of the minimum $S$ caused by the low $\gamma \sim
1.5$ at $\theta \ll 10\arcsec$ means that even the most
sensitive SKA images will not be confusion limited at beamwidths
$\theta \gtrsim 1\arcsec$.  This is fortunate because the median
angular size of faint star-forming galaxies is $\langle \phi
\rangle = 0\,\farcs3 \pm 0\,\farcs1$ with an rms scatter $\sigma_\phi
\lesssim 0\,\farcs3$ \citep{cotton18}, and sources with $\phi
\gtrsim \theta$ are more difficult to detect because their peak
flux densities are reduced by factors $\gtrsim 2$.  Although the
\citet{loi19} 1.4~GHz simulation of the radio sky for the SKA and
its precursors also
predicts $\gamma \lesssim 1.5$ below $S = 10~\mu\mathrm{Jy}$, their
estimated rms confusion $\sigma_\mathrm{mJy/bm} = (0.237 \pm 0.001)
(\nu_\mathrm{GHz})^{-0.8} (\theta)^{2.149 \pm 0.001}$ (the
dotted line in Figure~\ref{fig:sigconf2} shows their
$5\sigma_\mathrm{mJy/bm}$) does not take the lower $\gamma$ into
account and overpredicts the rms confusion at nJy flux densities.

Statistical counts using the $P(D)$ distribution can usefully reach
much lower $\beta$ values and hence much lower flux densities.  For
point sources randomly placed on the sky, the Poisson probability that
any beam solid angle will contain \emph{no} sources stronger than $S$
is ${P_\mathrm{P} = \exp (-1/\beta)}$.  Half of all beam solid angles
(${P_\mathrm{P} = 0.5}$) must satisfy $\beta = 1/\ln(2) \approx 1.44$,
and the flux density of the strongest source in them, $\langle S
\rangle$, is the solution of
\begin{equation}
  N(>\langle S \rangle) = \frac {\ln 2}{\Omega_\mathrm{b}}
  = \frac{1}{\pi}
  \Biggl( \frac {2 \ln 2}{\theta} \Biggr)^2~.
\label{eqn:softheta}
\end{equation}

\begin{figure}
  \includegraphics[trim={1.5cm 9.8cm 4.50cm 7.0cm},
    clip, width=\columnwidth]{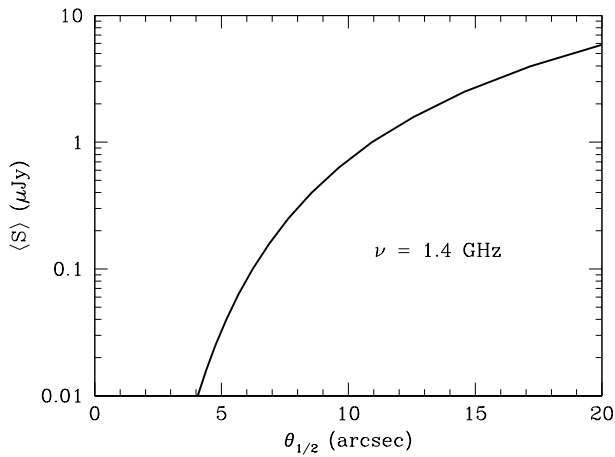}
  \caption{The solid curve based on Equation~\ref{eqn:softheta} shows
    the flux density $\langle S \rangle$ of the strongest source in
    half of the Gaussian beam areas of an image made with resolution
    $\theta$.}
\label{fig:sigconf3}
\end{figure}

For the $\theta = 7\,\farcs6$ FWHM DEEP2 beam and \citet{condon84}
model 1.4~GHz source count, Figure~\ref{fig:sigconf3} shows that
half of all beam solid angles contain no sources
stronger than $\langle S \rangle\approx 0.25\,\mu\mathrm{Jy}$.
Although the observed DEEP2 $P(D)$ distribution
(Figure~\ref{fig:pdall}) is broadened by $\sigma_\mathrm{n} \approx
0.55 \,\mu\mathrm{Jy~beam}^{-1}$ noise, it samples $\sim 1.2 \times
10^4$ independent effective beam areas $\Omega_\mathrm{e}$, so it
should be able to constrain the 1.4~GHz source count down to the $S
\approx 0.25\,\mu\mathrm{Jy}$ SKA1 goal for studying the
star-formation history of the universe \citep{prandoni15}.
At 1.4 GHz the ratio of the $S \sim 16~\mu\mathrm{Jy}$
detection limit $\beta_\mathrm{min} = 25$ for discrete sources to the
confusion sensitivity
limit $\langle S \rangle \approx 0.25 ~\mu\mathrm{Jy}$ is about 64!
Another advantage of $P(D)$ counts is that the low minimum $\beta \sim
1.44$ allows relatively large beamwidths, while the larger minimum
$\beta \approx 25$ for deep counts of individual sources requires
beamwidths small enough that large and difficult corrections for
partial source resolution become necessary \citep{owen18}.

\subsection{Direct Counts of DEEP2 Sources}
\label{sec:directcounts}

Figure~\ref{fig:sigconf2} indicates that the $\beta = 25$ limit
for reliably detecting individual sources with the DEEP2
$\theta_{1/2}  = 7\,\farcs 6$ beam is $S \approx 17\,\mu\mathrm{Jy}$ at
1.278~GHz ($\approx 16\,\mu\mathrm{Jy}$ at 1.4~GHz).
Below that limit, a significant fraction of sources will be confused
or obscured.  We produced preliminary counts of DEEP2 sources stronger than
$10\,\mu\mathrm{Jy}$ inside
the $\Omega \approx 1.026\mathrm{~deg}^2$ DEEP2 half-power circle.
To estimate confusion and obscuration corrections, we simulated a DEEP2 image,
counted sources in the simulated image, and compared those counts with the
actual counts used in the simulation.

For the simulation, point sources with approximately the correct
source count were randomly placed on the image and convolved with the
DEEP2 dirty beam. Then the simulated image was multiplied by the
primary attenuation, convolved with $0.55\,\mu\mathrm{Jy~beam}^{-1}$
rms noise, and CLEANed.  A source catalog was extracted from the
simulated image, and the source counts from this catalog were compared
with the input counts to derive count corrections.
These corrections are typically $\sim 10$\%,
and we estimate that their rms uncertainties are about half of the corrections themselves.
The brightness-weighted 1.278~GHz DEEP2 source counts
  $\log[S^2n(S)\mathrm{(Jy~sr}^{-1}\mathrm{)}]$ in 12 bins of logarithmic width
  $\Delta = 0.2$ in $\log(S)$ centered on
  $\langle \log[S\mathrm{(Jy)}] \rangle = -4.9, -4.7, \dots, -2.7$
  and their rms errors
  are listed in Table~\ref{tab:deep2counts}.

We converted the 1.278~GHz counts to 1.4~GHz via a median
spectral index $-0.7$, and these 1.4~GHz counts are
plotted as the black points in Figure~\ref{fig:counts}. The top
panel of Figure~\ref{fig:counts} presents the traditional static
Euclidean normalization $S^{5/2}n(S)$ and the bottom panel shows the
brightness-weighted normalization $S^2 n(S)$.  The red points are from
\citet{prandoni18}, the blue points are from \citet{hopkins03},
and the green points are from \citet{smolcic17}.  The DEEP2,
\citet{prandoni18}, and \citet{hopkins03} counts agree within the
errors.  The counts from \citet{smolcic17} are somewhat lower
in the range $-4 \lesssim \log[S\mathrm{(Jy)}] \lesssim -3$, perhaps
because some extended sources were partially resolved by their
$0\,\farcs75$ beam. Figure~\ref{fig:counts} also shows that the
DEEP2 direct count is slightly higher than the \citet{mitchell85} 68\%
confidence $P(D)$ box, possibly because DEEP2 is significantly more sensitive
(rms noise $\sigma_\mathrm{n} = 0.55 \,\mu\mathrm{Jy~beam}^{-1} \approx 7.2\mathrm{~mK}$)
to low-brightness star-forming galaxies.

%

\begin{table}
  \caption{DEEP2 1.278 GHz source counts}
  \label{tab:deep2counts}
\centering
  \begin{tabular}{c r c}
  \hline

  $\langle \log [S\mathrm{(Jy)}]\rangle$ &
    ${n}_{\rm bin}$ &
  $\log[S^2 n(S)\, \mathrm{(Jy~sr}^{-1}\mathrm{)}]$ \\
  \hline
 $-4.9$  & 5572 &  $2.682 \pm 0.031$ \\
 $-4.7$  & 4521 &  $2.754 \pm 0.031$ \\
 $-4.5$  & 3025 &  $2.762 \pm 0.032$ \\
 $-4.3$  & 2008 &  $2.770 \pm 0.032$ \\
 $-4.1$  & 1070 &  $2.740 \pm 0.033$ \\
 $-3.9$  &  576 &  $2.690 \pm 0.036$ \\
 $-3.7$  &  289 &  $2.578 \pm 0.040$ \\
 $-3.5$  &  139 &  $2.444 \pm 0.047$ \\
 $-3.3$  &   85 &  $2.450 \pm 0.055$ \\
 $-3.1$  &   49 &  $2.428 \pm 0.066$ \\
 $-2.9$  &   30 &  $2.411 \pm 0.080$ \\
 $-2.7$  &   24 &  $2.532 \pm 0.087$ \\
\hline
  \end{tabular}
  \end{table}

\section{DEEP2 and the Star Formation History of the Universe}
\label{sec:sfhu}

The 1.4~GHz continuum emission from a star-forming galaxy (SFG) is a
combination of synchrotron radiation from relativistic electrons
accelerated in the supernova remnants of short-lived ($\tau \lesssim
30 \mathrm{~Myr}$) massive stars and free-free radiation from thermal
electrons in H\textsc{ii} regions ionized by stars that are even more
massive and short-lived.  It is uniquely unbiased by dust or older
stars.  The tight and fairly linear FIR/radio correlation
\citep{condon91} indicates that the radio luminosity of a SFG is
directly proportional to its recent star-formation rate (SFR).  Thus
\begin{equation}
  \frac{{\rm SFR}(M> 0.1M_{\odot})}{M_{\odot}\,\rm yr^{-1}} =
  \kappa \left(\frac{L_{\rm 1.4\,GHz}}{\rm W\,Hz^{-1}}\right)~,
\label{eqn:sfrofl}
\end{equation}
where the constant of proportionality $\kappa$ is in the range $
(0.6\text{~--~}1.2) \times10^{-21}$ \citep{condon92, sullivan01,
  bell03, kennicutt09, murphy11} and depends on the unknown initial
mass function (IMF) below $M \sim 8 M_\odot$.  However, the only
consequence of the uncertainty in $\kappa$ is a global scaling in
the total mass of stars in the universe.  It does not alter the
determination of the evolutionary models of the SFRD through comparisons
of local radio luminosity functions with radio source counts. Therefore,
the local radio luminosity function and the DEEP2 confusion probability
distribution will constrain the luminosity and density evolution of SFGs
independent of dust, older stars, and SFR calibration errors.

The universe is homogeneous on large scales, so its spatially averaged
SFRD $\psi$ depends only on cosmic time or a proxy for time such as
redshift $z$.  It is usually written in units of $M_\odot
\mathrm{~yr}^{-1} \mathrm{~Mpc}^{-3}$.  Combining the 1.4~GHz local
luminosity function of SFGs and the source count of distant SFGs
yields an independent extinction-free means of measuring the star
formation history of the universe.  The main obstacle has been the
fact that SFGs are intrinsically weak radio sources, so tracing the
formation of most stars, and not just the tip of the iceberg in
ultraluminous starburst galaxies, requires counting sources fainter
than $S \sim 1~\mu\mathrm{Jy}$.

Our method for calculating the SFRD in the standard $\Lambda$CDM
universe with $\Omega_\mathrm{m} = 0.3$ and $H_0 = 70
\mathrm{~km~s}^{-1} \mathrm{~Mpc}^{-1}$ from radio data follows
Appendix C of \citet{condon18}.  Let $\rho (L_\nu \vert z)\, dL$ be
the number of sources with spectral luminosities $L_\nu$ to $L_\nu +
dL_\nu$ in comoving volume element $d V_\mathrm{C}$ and let
$\eta(S,z)\, dS\, dz$ be the number of sources per steradian with flux
densities $S$ to $S + dS$ in the redshift range $z$ to $z + dz$.  Then
\begin{equation}
  \eta(S, z)\, dS \,dz =  \rho (L_\nu|z)\, dL_\nu \,dV_\mathrm{C} ~,
\label{eqn:etasz}
\end{equation}
where the comoving volume element per sr is
\begin{equation}
dV_\mathrm{C} = \frac{D_\mathrm{C}^2 \,D_{\mathrm H_0}}{E(z)}dz~,
\end{equation}
$D_\mathrm{H_0} \equiv c / H_0$, and $E(z) \equiv H / H_0$ specifies
the expansion history of the universe.  For sources with spectral index
$\alpha$,
\begin{equation}
L_{\nu} = 4 \pi D_\mathrm{C}^2(1+z)^{1-\alpha}S~.
\end{equation}

Multiplying both sides of Equation~\ref{eqn:etasz} by $S^2$ converts
this redshift dependent differential source count into a
brightness-weighted source count
\begin{equation}
S^2\eta(S,z) = L_{\nu}^2
\rho(L_{\nu}|z) \Biggl[\frac {(1+z)^{\alpha-1}D_\mathrm{H_0}} {4 \pi E(z)} \Biggr] ~.
\end{equation}
Similarly, multiplying the spectral luminosity function $\rho(L_\nu \vert z)$ by
luminosity emphasizes the luminosity ranges contributing the most to
the spectral luminosity density $U(L_\nu|z) \equiv
L_{\nu}\rho(L_{\nu}|z)$. SFGs span several decades of luminosity, so
it is convenient to replace $U$ by the spectral luminosity density per
decade of luminosity
\begin{equation}
  U_\mathrm{dex}(L_\nu | z) = L\rho(L_\nu | z) \frac{dL}{d\log L} =
  L_\nu^2 \rho(L_\nu | z) \ln(10)~.
\end{equation}
The local energy density functions
$U_\mathrm{dex}(L_\mathrm{1.4~GHz}  \vert z = 0)$ for SFGs and
AGN-powered radio galaxies were separately derived from a spectroscopically
complete sample containing $9,517$
NVSS sources cross-identified with 2MASS galaxies \citep{condon19}. In terms
of $U_\mathrm{dex}$,
\begin{equation}
S^2 \eta(S,z) = U_{\rm
  dex}(L_{\nu}|z)\left[\frac{(1+z)^{\alpha-1}D_{\rm H_0}}{4 \pi
    \ln(10) E(z)}\right]~.
\end{equation}
The brightness-weighted source count is obtained by
integrating over redshift:
\begin{equation}
S^2 n(S) = \frac{D_{\rm H_0}}{4 \pi \ln(10)}\int_0^{\infty}U_{\rm
  dex}(L_{\nu}|z)\left[\frac{(1+z)^{\alpha-1}}{E(z)}\right]dz.
\label{eqn:s2n}
\end{equation}

Thus the evolving SFRD $\psi (z)$ can
be constrained by comparing the observed brightness-weighted 1.4~GHz
source count $S^2 n(S)$ with counts predicted by evolving the local
energy density function $U_{\rm dex}(L_\mathrm{1.4~GHz}|z=0)$ with redshift
$z$ using various evolutionary models
\citep[e.g.][]{madaudickinson14, hopkins06}.
For example, \citet{madaudickinson14} derived the SFRD evolutionary model
\begin{equation}
\frac{\psi(z)}{\psi(0)} = \frac{(1+z)^{2.7}}{1+[(1+z)/2.9]^{5.6}}.
\label{eqn:madau}
\end{equation}
by fitting available UV and infrared data.

\begin{figure}
\centering \includegraphics[trim = {2.3cm 8.9cm 3.5cm 13.5cm},clip,
  scale = 0.65]{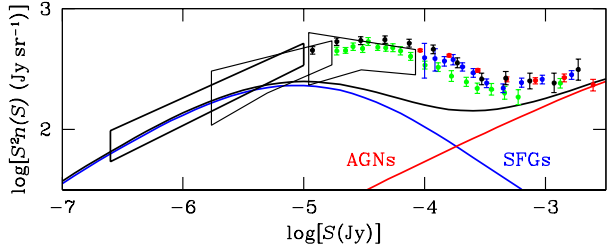}
\caption{The blue and red curves represent the brightness-weighted
  counts $S^2n(S)$ of
  SFGs \citep[from][]{madaudickinson14} and AGNs \citep[from][]{condon84},
  respectively.  Their sum is shown by the black curve, which lies
  significantly below the observed counts from Figure~\ref{fig:counts} in
  the range $-5 \lesssim \log[S\mathrm{(Jy)}] \lesssim -4$.}
  \label{fig:evolve}
\end{figure}

Pure luminosity evolution consistent with Equation~\ref{eqn:madau} for
SFGs predicts the brightness-weighted source counts shown in Figure
\ref{fig:evolve} by the blue curve.  The red curve is an estimate of
the AGN contribution \citep{condon84}.  The black curve is their sum,
and it is significantly lower than the actual 1.4~GHz source count in
the range $-5 \lesssim \log[S\mathrm{(Jy)}] \lesssim -4$.
The DEEP2 source count extends $\sim 7 \times$ deeper than the \citet{condon12}
source count (Figure~\ref{fig:evolve}), constraining for the first time
star formation in all galaxies with SFRs as low as $5\,M_\odot
\mathrm{yr}^{-1}$ at ``cosmic noon" ($z \approx 2$), not just the rare
starbursts with $\mathrm{SFR} > 35\,M_\odot \mathrm{yr}^{-1}$.  DEEP2
also reduces the statistical uncertainty in $S^2 n(S)$ by a factor of
four because it samples a much larger solid angle of sky.

While this simple preliminary analysis suggests that SFGs evolve more
strongly, it uses a power-law approximation for the faint-source
$P(D)$ counts and assumes a perfectly Gaussian beam.  To fully exploit
our DEEP2 image, we are now developing numerical simulations of
synthetic images populated with sources having arbitrary counts and
beams that take the limitations of CLEAN into account (Matthews et
al., in prep).

\acknowledgements
The MeerKAT telescope is operated by the South African Radio Astronomy
Observatory, which is a facility of the National Research
Foundation, an agency of the Department of Science and Innovation.
The National Radio Astronomy Observatory is a facility of the National
Science Foundation operated by Associated Universities, Inc.  This
material is based upon work supported by the National Science
Foundation Graduate Research Fellowship under Grant No. DDGE-1315231.
We thank our anonymous referee for a detailed and constructive review.

\facility{MeerKAT}



\bibliographystyle{aasjournal}

\bibliography{DEEP2.bib}




\end{document}